\documentclass[11pt]{article}
\usepackage{amssymb}
\usepackage{amsmath}
\usepackage{amsfonts}
 \usepackage{pifont}
\usepackage{amsmath,amsfonts,amssymb,theorem,graphicx,listings,txfonts}
\usepackage{graphics}
 \usepackage{caption2}
\usepackage{flafter}
 \usepackage{indentfirst}
\usepackage{epsfig}
  \usepackage{mathrsfs}
\usepackage{subfigure}
 \usepackage{cite}
\newtheorem{theorem}{Theorem}[section]

\newtheorem{proposition}[theorem]{Proposition}

\newtheorem{definition}[theorem]{Definition}

\theoremstyle{example}
\newtheorem{example}[theorem]{Example}
\theoremstyle{programme}

\theoremstyle{property}

\theoremstyle{problem}

\title{The construction of characteristic matrixes of dynamic coverings using an incremental approach}

\author
{Guangming Lang$^{a}$ \hspace{1cm} Qingguo Li$^{a}$
\thanks{Corresponding author.\quad Tel./fax: +86 731 8822855,
liqingguoli@yahoo.com.cn
\newline\mbox{}\hspace{0.55cm}
E-mail address: langguangming1984@126.com(M.Lang),
liqingguoli@yahoo.com.cn(G.Li),lankun.guo@gmail.com(K.Guo),
angchunyong2006@sina.com(C.Y. Wang).}\hspace{1cm}
Lankun Guo$^{b}$ \hspace{1cm} Chunyong Wang$^{a}$\\
\small {$^{a}$ College of Mathematics and Econometrics, Hunan University}\\
\small {Changsha, Hunan 410082, P.R. China}\\
\small {$^{b}$ College of Computer and Communicaton, Hunan University}\\
\small {Changsha, Hunan 410082, P.R. China}}
\date{}

\begin{document}
\maketitle \baselineskip=17pt
\begin{center}
\begin{quote}
{{\bf Abstract.} The covering approximation space evolves in time
due to the explosion of the information, and the characteristic
matrixes of coverings viewed as an effective approach to
approximating the concept should update with time for knowledge
discovery. This paper further investigates the construction of
characteristic matrixes without running the matrix acquisition
algorithm repeatedly. First, we present two approaches to computing
the characteristic matrixes of the covering with lower time
complexity. Then, we investigate the construction of the
characteristic matrixes of the dynamic covering using the
incremental approach. We mainly address the characteristic matrix
updating from three aspects: the variations of elements in the
covering, the immigration and emigration of objects and the changes
of attribute values. Afterwards, several illustrative examples are
employed to show that the proposed approach can effectively compute
the characteristic matrixes of the dynamic covering for
approximations of concepts.

\\}
\end{quote}
\end{center}
\renewcommand{\thesection}{\arabic{section}}

\section{Introduction}

Covering-based rough set theory\cite{Zakowski1}, as a powerful
mathematical tool for studying information systems, has attracted a
lot of attention in recent years. Actually, the covering rough sets
is regarded as a meaningful extension of Pawlak's
model\cite{Pawlak1,Pawlak2,Pawlak3,Pawlak4} to deal with more
complex data sets. Nowadays
it\cite{Chen2,Deng1,Diker1,Feng1,Li2,Wu1,Xiang1,Xu1,Tsang1,Yao1,Yao2,
Yang1,Yang2,Yun1,Zhang2,Zhang3,Wang2} has been successfully applied
to pattern recognition, machine learning and environmental science
because of its approximation ability.

Constructing the approximations of concepts is one major work of the
covering-based rough set theory by using reasonable approximation
operators. On one hand, the
literatures\cite{Yang1,Yang2,Zhu1,Zhu2,Zhu3,Zhu4,Zhu5,Zhu6,Zhu7,Zhu8,Zhu9}
have already provided several models of covering-based rough sets
and multiple fuzzy rough set models based on fuzzy coverings. On the
other hand, researchers studied the basic properties of the proposed
models of covering-based rough sets. For example,
Mordeson\cite{Mordeson1} discussed the algebraic structural
properties of certain subsets for a type of covering-based rough
sets. Wang et al.\cite{Wang2} investigated the data compression of
the covering information system. Yang et al.\cite{Yang1} unified the
reduction of different types of covering generalized rough sets. Zhu
and Wang\cite{Zhu1,Zhu2,Zhu3,Zhu4,Zhu5,Zhu6,Zhu7,Zhu8} studied five
types of covering-based rough sets systematically. Naturally, it
motivates us to further study the covering-based rough set theory.

Recently, Wang et al.\cite{Wang1} represented and axiomatized three
types of covering approximation operators by using two types of
characteristic matrixes. In other words, the computation of the
approximation of a set is transformed into the product of the
characteristic matrix of the covering and the characteristic
function of the set. The results presents a new view to discuss the
covering-based rough sets by borrowing extensively from boolean
matrixes. Actually, one major work is to construct the two types of
characteristic matrixes of the covering in the process of computing
the approximations of concepts. But Wang et al.\cite{Wang1} paid
little attention to the approach to constructing the characteristic
matrixes, and it is of interest to introduce an effective approach
to computing the characteristic matrixes with low time complexity.
On the other hand, the covering approximation space varies with time
due to the characteristics of data collection in practice, and the
non-incremental approach to constructing the approximations of
concepts in the dynamic covering approximation space is often very
costly or even intractable. As we know, some
scholars\cite{Chen1,Li1,Liu1,Liu2,Zhang1} studied attribute
reductions of the dynamic information systems. Therefore, it is
interesting to apply an incremental updating scheme to maintain the
approximations of sets dynamically and avoid unnecessary
computations by utilizing the approximations in the original
covering approximation space.

The purpose of this paper is to further study the approximations of
concepts using the characteristic matrixes. First, we introduce two
approaches to approximating the concept using the characteristic
matrixes of the covering. More concretely, we can compress the
covering approximation space into a small-scale one under the
condition of the consistent function and obtain the approximation of
the concept by computing the approximation of its image. Thus the
characteristic matrixes of the covering can be transformed into a
small one, and it can reduce the time complexity of constructing the
approximation of the concept. Subsequently, we present another
approach to computing the characteristic matrixes of the covering by
constructing the matrix representation of each element in the
covering, and it is useful for the construction of the
characteristic matrixes of the dynamic covering. Especially, we can
apply the two approaches to constructing the approximation of the
concept simultaneously. Second, we show that how to get the
characteristic matrixes of the dynamic covering by using an
incremental approach. We mainly focus on five types of dynamic
coverings: adding elements into the covering and deleting the
elements of the covering, the immigration and emigration of the
object sets, and revising the attribute values of some objects. We
investigate the relationship between the characteristic matrixes of
the original and dynamic coverings. Several examples are employed to
illustrate that how to update the characteristic matrixes of the
dynamic coverings by utilizing an incremental approach.

The rest of this paper is organized as follows: Section 2 briefly
reviews the basic concepts of the covering information systems. In
Section 3, we present two approaches to computing the characteristic
matrixes of the covering. Section 4 is devoted to constructing the
type-1 and type-2 characteristic matrixes of the dynamic covering by
utilizing the incremental approach. We conclude the paper in Section
5.

\section{Preliminaries}

In this section, we review some concepts of the covering and
characteristic matrixes. In addition, we investigate the basic
properties of the characteristic matrixes.

\begin{definition}\cite{Zakowski1}
Let $U$ be a finite universe of discourse, and $\mathscr{C}$ a
family of subsets of $U$. If $\mathscr{C}$ satisfies the conditions:

$(1)$ none of elements of $\mathscr{C}$ is empty;

$(2)$ $\bigcup\{C|C\in \mathscr{C}\}=U$, then $\mathscr{C}$ is
called a covering of $U$.
\end{definition}

It is obvious that the concept of the covering is an extension of
the partition. In addition, $(U,\mathscr{C})$ is called a covering
approximation space if $\mathscr{C}$ is a covering of the universe
$U$.

To compress the covering approximation space, Wang et
al.\cite{Wang2} provided the concept of consistent functions with
respect to the covering.

\begin{definition}\cite{Wang2}
Let $f$ be a mapping from $U_{1}$ to $U_{2}$,
$\mathscr{C}$=$\{C_{1}, C_{2},...,C_{N}\}$    a covering of $U_{1}$,
$N(x) $=$\bigcap\{C_{i} | x\in C_{i}  \text{ and } C_{i} \in
\mathscr{C} \}$, and $[x]_{f}=\{y\in U_{1}| f(x)=f(y)\}$.  If
$[x]_{f}\subseteq N(x)$ for any $x\in U_{1}$, then $f$ is called a
consistent function with respect to $\mathscr{C}$.
\end{definition}

Based on Definition 2.2, Wang et al. constructed a homomorphism
between a complex massive covering information system and a
relatively small-scale covering information system, and the
homomorphism provides the foundation for the communication between
covering information systems.

Recently, Wang et al.\cite{Wang1} introduced the concepts of
characteristic matrixes for constructing the approximations of
concepts.

\begin{definition}\cite{Wang1}
Let $U=\{x_{1},x_{2},...,x_{n}\}$ be a finite universe,
$\mathscr{C}=\{C_{1},C_{2},...,C_{m}\}$ a family of subsets of $U$,
and $M_{\mathscr{C}}=(c_{ij})_{n\times m}$, where $c_{ij}=\left\{
\begin{array}{ccc}
1,&{\rm}& x_{i}\in C_{j};\\
0,&{\rm}& x_{i}\notin C_{j}.
\end{array}
\right. $ Then $M_{\mathscr{C}}$ is called a matrix representation
of $\mathscr{C}$.
\end{definition}

In the sense of Definition 2.3, if $\mathscr{C}$ is a covering of
the universe $U$, then $M_{\mathscr{C}}\cdot M_{\mathscr{C}}^{T}$ is
called the type-1 characteristic matrix of $\mathscr{C}$, denoted as
$\Gamma(\mathscr{C})$, where $M_{\mathscr{C}}\cdot
M_{\mathscr{C}}^{T}$ is the boolean product of $M_{\mathscr{C}}$ and
its transpose $ M_{\mathscr{C}}^{T}$. Furthermore,
$\Gamma(\mathscr{C})$ is the relational matrix of the relation
induced by indiscernible neighborhoods of the covering, denoted as
$R_{\mathscr{C}}$. Especially, we have that $(x,y)\in
R_{\mathscr{C}}$ iff $y\in I(x)=\{K\in \mathscr{C}|x\in K\}$ for all
$x,y\in U$.

\begin{proposition}
Let $U=\{x_{1},x_{2},...,x_{n}\}$ be a finite universe,
$\mathscr{C}=\{C_{1},C_{2},...,C_{m}\}$ a family of subsets of $U$,
and $M_{C}\cdot M_{C}^{T}=(c_{ij})_{n\times n}$ the type-1
characteristic matrix of $C\in \mathscr{C}$. Then we have that
$c_{ii}=1$ iff $x_{i}\in C$.
\end{proposition}

\noindent\textbf{Proof.} Suppose that
$M_{C}=[a_{1},a_{2},...,a_{n}]^{T}$ is the matrix representation of
$C\in \mathscr{C}$. In the sense of Definition 2.3, we have that
$a_{i}=1$ if $x_{i}\in C$. It follows that $a_{i}\wedge a_{i}=1$.
Thus, $c_{ii}=1$.

By Definition 2.3, we have that $c_{ij}=a_{i}\wedge a_{j}$. It
implies that $a_{i}=1$. Therefore, $x_{i}\in C$. $\Box$

\begin{proposition}
Let $U=\{x_{1},x_{2},...,x_{n}\}$ be a finite universe,
$\mathscr{C}=\{C_{1},C_{2},...,C_{m}\}$ a family of subsets of $U$,
and $M_{A}\cdot M_{A}^{T}=(a_{ij})_{n\times n}$, $M_{B}\cdot
M_{B}^{T}=(b_{ij})_{n\times n}$, $M_{C}\cdot
M_{C}^{T}=(c_{ij})_{n\times n}$ the type-1 characteristic matrixes
of $A, B, C\in \mathscr{C}$, respectively. Then we have that
$c_{ii}=a_{ii}\vee b_{ii}$ iff $C$ is the union of $A$ and $B$.
\end{proposition}

\noindent\textbf{Proof.}  By Proposition 2.4, if $c_{ii}=1$, then we
have that $a_{ii}=1$ or $b_{ii}=1$. It implies that $x_{i}\in A$ or
$x_{i}\in B$ if $x_{i}\in C$. Furthermore, if $c_{ii}=0$, then
$a_{ii}=0$ and $b_{ii}=0$. In other words, $x_{i}\notin A$ and
$x_{i}\notin B$ if $x_{i}\notin C$. Thus $C$ is the union of $A$ and
$B$.

If $C$ is the union of $A$ and $B$, then that $x_{i}\in C$ implies
that $x_{i}\in A$ or $x_{i}\in B$. It follows that $c_{ii}=1$ if
$a_{ii}=1$ or $b_{ii}=1$. Thus $c_{ii}=a_{ii}\vee b_{ii}$. Moreover,
that $x_{i}\notin C$ implies that $x_{i}\notin A$ and  $x_{i}\notin
B$. Then we have that $a_{ii}=0$, $b_{ii}=0$ and $c_{ii}=0$. It
follows that $c_{ii}=a_{ii}\vee b_{ii}$. Therefore,
$c_{ii}=a_{ii}\vee b_{ii}$ if $C$ is the union of $A$ and $B$.
$\Box$

By Proposition 2.5, we have that $c_{ii}=a_{ii}\vee b_{ii}$ iff $C$
is the union of $A$ and $B$. An example is employed to illustrate
that $M_{C}\cdot M_{C}^{T}=(a_{ij})_{n\times n}=M_{A}\cdot
M_{A}^{T}\vee M_{B}\cdot M_{B}^{T}$ does not necessarily hold if $C$
is the union of $A$ and $B$.

\begin{example}
Let $U=\{x_{1},x_{2},x_{3},x_{4}\}$, $A=\{x_{1},x_{2}\}$,
$B=\{x_{1},x_{4}\}$ and $C=\{x_{1},x_{2},x_{4}\}$. By Definition
2.3, we have that $M_{A}\cdot M_{A}^{T}=\left[
\begin{array}{cccc}
1 & 1 & 0 & 0 \\
1 & 1 & 0 & 0 \\
0 & 0 & 0 & 0 \\
0 & 0 & 0 & 0 \\
\end{array}
\right],$
 $M_{B}\cdot M_{B}^{T}=\left[
\begin{array}{cccc}
1 & 0 & 0 & 1 \\
0 & 0 & 0 & 0 \\
0 & 0 & 0 & 0 \\
1 & 0 & 0 & 1 \\
\end{array}
\right],$ and $M_{C}\cdot M_{C}^{T}=\left[
\begin{array}{cccc}
1 & 1 & 0 & 1 \\
1 & 0 & 0 & 1 \\
0 & 0 & 0 & 0 \\
1 & 1 & 0 & 1 \\
\end{array}
\right].$ It is obvious that $C=A\cup B$. But $M_{C}\cdot
M_{C}^{T}\neq M_{A}\cdot M_{A}^{T}\vee M_{B}\cdot M_{B}^{T}$.
\end{example}

\begin{proposition}
Let $U=\{x_{1},x_{2},...,x_{n}\}$ be a finite universe,
$\mathscr{C}=\{C_{1},C_{2},...,C_{m}\}$ a family of subsets of $U$,
and $M_{A}\cdot M_{A}^{T}=(a_{ij})_{n\times n}$, $M_{C}\cdot
M_{B}^{T}=(b_{ij})_{n\times n}$, $M_{C}\cdot
M_{C}^{T}=(c_{ij})_{n\times n}$ the type-1 characteristic matrixes
of $A, B$ and $C\in \mathscr{C}$, respectively. Then

$(1)$ $C$ is the intersection of $A$ and $B$ iff
$c_{ii}=a_{ii}\wedge b_{ii}$;

$(2)$ $C$ is the intersection of $A$ and $B$ iff $M_{C}\cdot
M_{C}^{T}=M_{A}\cdot M_{A}^{T}\wedge M_{B}\cdot M_{B}^{T}$.
\end{proposition}

\noindent\textbf{Proof.} $(1)$ The proof is similar to that of
Proposition 2.5.

$(2)$ By Proposition 2.7 $(1)$, the proof is straightforward. $\Box$

\begin{definition}\cite{Wang1}
Let $A=(a_{ij})_{n\times m}$ and $B=(b_{ij})_{m\times p}$ be two
boolean matrixes, and $C=A\odot B=(c_{ij})_{n\times p}$. Then
$c_{ij}$ is defined as
$$c_{ij}=\bigwedge^{m}_{k=1}(b_{kj}-a_{ik}+1).$$
\end{definition}

In the sense of Definition 2.8, if $\mathscr{C}$ is a covering of
the universe of $U$, then $M_{\mathscr{C}}\odot M_{\mathscr{C}}^{T}$
is called the type-2 characteristic matrix of $\mathscr{C}$, denoted
as $\prod(\mathscr{C})$. Furthermore, $\prod(\mathscr{C})$ is the
relational matrix of the relation induced by neighborhoods of the
covering, denoted as $R(\mathscr{C})$. Specially, we have that
$(x,y)\in R(\mathscr{C})$ iff $y\in N(x)$ for all $x,y\in U$.

\begin{definition}\cite{Wang1}
Let $U=\{x_{1},x_{2},...,x_{n}\}$ be a finite universe, and
$\mathscr{C}=\{C_{1},C_{2},...,C_{m}\}$ a covering of $U$. For any
$X\subseteq U$, the second, fifth and sixth upper and lower
approximations of $X$ with respect to $\mathscr{C}$, respectively,
are defined as follows:

$(1)$ $SH_{\mathscr{C}}(X)=\bigcup\{C\in\mathscr{C}|C\cap X\neq
\emptyset\}$, $SL_{\mathscr{C}}(X)=[SH_{\mathscr{C}}(X^{c})]^{c}$;

$(2)$ $IH_{\mathscr{C}}(X)=\bigcup\{x\in U|N(x)\cap X\neq
\emptyset\}$, $IL_{\mathscr{C}}(X)=\bigcup\{x\in U|N(x)\subseteq
X\}$;

$(3)$ $XH_{\mathscr{C}}(X)=\bigcup\{N(x)\in U|N(x)\cap X\neq
\emptyset\}$, $XL_{\mathscr{C}}(X)=\bigcup\{N(x)\in U|N(x)\subseteq
X\}$, where $N(x)=\bigcap\{K\in\mathscr{C}|x\in K\}$.
\end{definition}

Throughout the paper, we delete $\mathscr{C}$ in the description of
the lower and upper approximation operators for simplicity. By using
the type-1 and type-2 characteristic matrixes, Wang et al.
represented equivalently and axiomatized three important types of
covering approximation operators.

\begin{definition}\cite{Wang1}
Let $U=\{x_{1},x_{2},...,x_{n}\}$ be a finite universe, and
$\mathscr{C}=\{C_{1},C_{2},...,C_{m}\}$ a covering of $U$. Then

$(1)$ $\mathcal {X}_{SH(X)}=\Gamma(\mathscr{C})\cdot \mathcal
{X}_{X}$, $\mathcal {X}_{SL(X)}=\Gamma(\mathscr{C})\odot \mathcal
{X}_{X}$;

$(2)$ $\mathcal {X}_{IH(X)}=\prod(\mathscr{C})\cdot \mathcal
{X}_{X}$, $\mathcal {X}_{IL(X)}=\prod(\mathscr{C})\odot \mathcal
{X}_{X}$;

$(3)$ $\mathcal {X}_{XH(X)}=\prod(\mathscr{C})^{T}\cdot
\prod(\mathscr{C})\cdot \mathcal {X}_{X}$, $\mathcal
{X}_{XL(X)}=\prod(\mathscr{C})^{T}\cdot \prod(\mathscr{C})\odot
\mathcal {X}_{X}$, where $\mathcal {X}_{X}$ denotes the
characteristic function of $X$ in $U$. In other words, for any $y\in
U$, $\mathcal {X}_{X}(x)=1$ iff $x\in X$.
\end{definition}

\begin{proposition}
Let $U=\{x_{1},x_{2},...,x_{n}\}$ be a finite universe, and
$\mathscr{C}=\{C_{1},C_{2},...,C_{m}\}$ a covering of $U$. Then
$\prod(\mathscr{C})\leq \Gamma(\mathscr{C})$ and
$\prod(\mathscr{C})^{T}\cdot \prod(\mathscr{C})=\prod(\mathscr{C})$.
\end{proposition}

\noindent\textbf{Proof.} For all $x,y\in U$, if we have that $y\in
N(x)$, then it implies that $y\in I(x)$. But the converse does not
hold necessarily. It follows that $R_{\mathscr{C}}\subseteq
R(\mathscr{C})$. Thus $\prod(\mathscr{C})\leq \Gamma(\mathscr{C})$.

Since $IH_{\mathscr{C}}(X)=XH_{\mathscr{C}}(X)$, we obtain that
$\mathcal {X}_{IH(X)}=\mathcal {X}_{XH(X)}.$ Therefore,
$\prod(\mathscr{C})^{T}\cdot \prod(\mathscr{C})=\prod(\mathscr{C})$.
$\Box$

\section{Two approaches to constructing approximations of concepts using characteristic matrixes}

In this section, we introduce two approaches to constructing the
approximation of the concept by using the characteristic matrixes,
and the proposed approaches can be applied to the dynamic covering
approximation space.

Suppose that $(U, \mathscr{C})$ is a covering approximation space,
where $U=\{x_{1},x_{2},...,x_{n}\}$ is a finite universe, and
$\mathscr{C}=\{C_{1},C_{2},...,C_{m}\}$ is a covering of $U$. In the
sense of Definitions 2.3 and 2.8, we see that $\prod(\mathscr{C})$
and $\Gamma(\mathscr{C})$ are $n\times n$ matrixes if the number of
the objects of the universe is $n$. Obviously, the time complexity
of computing the approximation of the concept by using the
approach\cite{Wang1} is high if there are a large number of objects
in the universe. To solve this issue, we introduce an approach to
constructing the approximation of the concept by using the
consistent function. There are two steps for computing the
approximation of the concept. First, we compress the covering
approximation space $(U,\mathscr{C})$ into a relative small scale
one $(f(U),f(\mathscr{C}))$ by using the consistent function $f$.
Then we obtain the approximation $f^{-1}(Y)$ of $X$ in
$(U,\mathscr{C})$ by constructing the approximation $Y$ of $f(X)$ in
the $(f(U),f(\mathscr{C}))$.

An example is employed to illustrate the process of computing
approximations of concepts by using the proposed approach.

\begin{example}
Let $U=\{x_{1},x_{2},x_{3},x_{4},x_{5},x_{6}\}$,
$\mathscr{C}=\{C_{1},C_{2},C_{3}\}$, $C_{1}=\{x_{1},x_{2}\}$,
$C_{2}=\{x_{3},x_{4},x_{5},x_{6}\}$,
$C_{3}=\{x_{1},x_{2},x_{5},x_{6}\}$, and
$X=\{x_{1},x_{2},x_{3},x_{4}\}$. Then
$$\Gamma(\mathscr{C})=M_{\mathscr{C}}\cdot M_{\mathscr{C}}^{T}=
\left[
\begin{array}{c c c}
1 & 0 & 1\\
1 & 0 & 1\\
0 & 1 & 0\\
0 & 1 & 0\\
0 & 1 & 1\\
0 & 1 & 1\\
\end{array}
\right]\cdot\left[
\begin{array}{cccccc}
1 & 1 & 0 & 0 & 0 & 0\\
0 & 0 & 1 & 1 & 1 & 1\\
1 & 1 & 0 & 0 & 1 & 1\\
\end{array}
\right]=\left[
\begin{array}{cccccc}
1 & 1 & 0 & 0 & 1 & 1\\
1 & 1 & 0 & 0 & 1 & 1\\
0 & 0 & 1 & 1 & 1 & 1\\
0 & 0 & 1 & 1 & 1 & 1\\
1 & 1 & 1 & 1 & 1 & 1\\
1 & 1 & 1 & 1 & 1 & 1\\
\end{array}
\right].
$$

First, we get the upper and lower approximations of $X$ by computing
$$\mathcal
{X}_{SH(X)}
=\Gamma(\mathscr{C})\cdot \mathcal {X}_{X}
=\left[
\begin{array}{cccccc}
1 & 1 & 0 & 0 & 1 & 1\\
1 & 1 & 0 & 0 & 1 & 1\\
0 & 0 & 1 & 1 & 1 & 1\\
0 & 0 & 1 & 1 & 1 & 1\\
1 & 1 & 1 & 1 & 1 & 1\\
1 & 1 & 1 & 1 & 1 & 1\\
\end{array}
\right]\cdot \left[\begin{array}{c}
1 \\
1 \\
1 \\
1 \\
0 \\
0 \\
\end{array}
\right]
=\left[\begin{array}{c}
1 \\
1 \\
1 \\
1 \\
1 \\
1 \\
\end{array}
\right]$$ and

$$\mathcal
{X}_{SL(X)}
=\Gamma(\mathscr{C})\odot \mathcal {X}_{X}
=\left[
\begin{array}{cccccc}
1 & 1 & 0 & 0 & 1 & 1\\
1 & 1 & 0 & 0 & 1 & 1\\
0 & 0 & 1 & 1 & 1 & 1\\
0 & 0 & 1 & 1 & 1 & 1\\
1 & 1 & 1 & 1 & 1 & 1\\
1 & 1 & 1 & 1 & 1 & 1\\
\end{array}
\right]\odot \left[\begin{array}{c}
1 \\
1 \\
1 \\
1 \\
0 \\
0 \\
\end{array}
\right]
=\left[\begin{array}{c}
1 \\
1 \\
0 \\
0 \\
0 \\
0 \\
\end{array}
\right].$$

Second, we construct the consistent function $f$ from $U$ to $V$ as
follows:
$$f(x_{1})=f(x_{2})=y_{1}, f(x_{3})=f(x_{4})=y_{2},
f(x_{5})=f(x_{6})=y_{3},$$ and get $(V, f(\mathscr{C}))$, where
$V=\{y_{1},y_{2},y_{3}\}$,
$f(\mathscr{C})=\{f(C_{1}),f(C_{2}),f(C_{3})\}$,
$f(C_{1})=\{y_{1}\}$, $f(C_{2})=\{y_{2},y_{3}\}$, and
$f(C_{3})=\{y_{1},y_{3}\}$. Thus we can compress $X$ into
$f(X)=\{y_{1},y_{2}\}$ and get the lower and upper approximations of
$f(X)$ by computing $\mathcal
{X}_{SH(f(X))}=\Gamma(f(\mathscr{C}))\cdot \mathcal {X}_{f(X)}$ and
$\mathcal {X}_{SL(f(X))}=\Gamma(f(\mathscr{C}))\odot \mathcal
{X}_{f(X)}$ as follows:
\begin{eqnarray*}
\Gamma(f(\mathscr{C}))&=&M_{f(\mathscr{C})}\cdot
M_{f(\mathscr{C})}^{T}= \left[
\begin{array}{c c c}
1 & 0 & 1\\
0 & 1 & 0\\
0 & 1 & 1\\
\end{array}
\right]\cdot\left[
\begin{array}{cccccc}
1 & 0  & 0 \\
0 & 1  & 1 \\
1 & 0  & 1 \\
\end{array}
\right]=\left[
\begin{array}{cccccc}
1  & 0  & 1 \\
0  & 1  & 1 \\
1  & 1  & 1 \\
\end{array}
\right]; \\\mathcal {X}_{SH(f(X))} &=&\Gamma(f(\mathscr{C}))\cdot
\mathcal {X}_{f(X)} =\left[
\begin{array}{cccccc}
1  & 0  & 1 \\
0  & 1  & 1 \\
1  & 1  & 1 \\
\end{array}
\right]\cdot \left[\begin{array}{c}
1 \\
1 \\
0 \\
\end{array}
\right] =\left[\begin{array}{c}
1 \\
1 \\
1 \\
\end{array}
\right];\\
\mathcal {X}_{SL(f(X))} &=&\Gamma(\mathscr{C})\odot \mathcal
{X}_{f(X)} =\left[
\begin{array}{cccccc}
1  & 0  & 1 \\
0  & 1  & 1 \\
1  & 1  & 1 \\
\end{array}
\right]\odot \left[\begin{array}{c}
1 \\
1 \\
0 \\
\end{array}
\right] =\left[\begin{array}{c}
1 \\
0 \\
0 \\
\end{array}
\right].\end{eqnarray*}

It is obvious that $SH(f(X))=\{y_{1},y_{2},y_{3}\}$ and
$SL(f(X))=\{y_{1}\}$. Therefore,
$SH(X)=f^{-1}(SH(f(X)))=\{x_{1},x_{2},x_{3},x_{4},x_{5},x_{6}\}$ and
$SL(X)=f^{-1}(SL(f(X)))=\{x_{1}, x_{2}\}$.
\end{example}

Subsequently, we introduce another approach to constructing the
approximation of the concept by using the characteristic matrixes.
In the sense of Definition 2.3, we have that
$M_{C}=(d_{i1})_{n\times 1}$ for any $C\in \mathscr{C}$, where
$d_{i1}=\left\{
\begin{array}{ccc}
1,&{\rm}& x_{i}\in C;\\
0,&{\rm}& x_{i}\notin C
\end{array}
,\right. $ and $M_{C}$ is called the matrix representation of $C$.
Actually, we can obtain $M_{C}\cdot M_{C}^{T}=(c_{ij})_{n\times n}$
by identifying the value of $d_{i1}$. Concretely, we have that
$c_{ik}=d_{k1}$ $(1\leq k\leq n)$ if $d_{i1}=1$. Otherwise, we have
that $c_{ik}=0$ for $1\leq k\leq n$ if $d_{i1}=0$. Thereby, we only
need to identify $d_{i1}=1$ or $d_{i1}=0$.

We employ the following example to show the process of computing
$M_{C}\cdot M_{C}^{T}$ with the proposed approach.

\begin{example}
Let $U=\{x_{1},x_{2},x_{3},x_{4}\}$, $C=\{x_{1},x_{4}\}$, and
$M_{C}=(d_{ij})_{4\times 1}=[1~ 0 ~0 ~1]^{T}$. Then
$$M_{C}\cdot M_{C}^{T}=
\left[
\begin{array}{c}
1 \\
0 \\
0 \\
1 \\
\end{array}
\right]\cdot\left[
\begin{array}{cccc}
1 & 0 & 0 & 1 \\
\end{array}
\right]=\left[
\begin{array}{cccc}
c_{11} & c_{12} & c_{13} & c_{14} \\
c_{21} & c_{22} & c_{23} & c_{24} \\
c_{31} & c_{32} & c_{33} & c_{34} \\
c_{41} & c_{42} & c_{43} & c_{44} \\
\end{array}
\right]=\left[
\begin{array}{cccc}
1 & 0 & 0 & 1 \\
0 & 0 & 0 & 0 \\
0 & 0 & 0 & 0 \\
1 & 0 & 0 & 1 \\
\end{array}
\right].
$$
\end{example}

From the above results, we see that $d_{11}=d_{41}=1$ and
$d_{21}=d_{31}=0$. It is obvious that $c_{1j}=c_{4j}=d_{j1}$ and
$c_{2j}=c_{3j}=0$ $(1\leq j\leq 4)$. Therefore, we can get
$M_{C}\cdot M_{C}^{T}$ without computing $c_{ij}$ $(1\leq i, j\leq
4)$, and the time complexity of computing $M_{C}\cdot M_{C}^{T}$ can
be reduced greatly.

Then we investigate the construction of the type-1 characteristic
matrix based on Example 3.2. For the covering approximation space
$(U,\mathscr{C})$, we obtain the matrix representations
$\{M_{C_{i}}\cdot M_{C_{i}}^{T}|C_{i}\in \mathscr{C}\}$ shown in
Table 1. Since Wang et al. have proved that $M_{\mathscr{C}}\cdot
M_{\mathscr{C}}^{T}=\bigvee_{C_{i}\in \mathscr{C}} M_{C_{i}}\cdot
M_{C_{i}}^{T}$, we can get $M_{\mathscr{C}}\cdot
M_{\mathscr{C}}^{T}$ by computing $M_{C_{i}}\cdot M_{C_{i}}^{T}$.
Especially, $\{M_{C_{i}}\cdot M_{C_{i}}^{T}|C_{i}\in \mathscr{C}\}$
are useful for computing the type-1 characteristic matrix of the
dynamic covering, which are illustrated in Section 4.

\begin{table}[htbp]
\caption{The type-1 characteristic matrix of each
$C_{i}\in\mathscr{C}$ and $\mathscr{C}$.} \tabcolsep0.2in
\begin{tabular}{c c c c c c c c}
\hline $U$  &$C_{1}$& $C_{2}$&.&.&. &$C_{m}$& $\mathscr{C}$\\
\hline
$M_{C}\cdot M_{C}^{T}$ & $M_{C_{1}}\cdot M_{C_{1}}^{T}$& $M_{C_{2}}\cdot M_{C_{2}}^{T}$&.&.&. &$M_{C_{m}}\cdot M_{C_{m}}^{T}$ & $M_{\mathscr{C}}\cdot M_{\mathscr{C}}^{T}$\\
\hline
\end{tabular}
\end{table}

The following example is employed to illustrate the computing of the
type-1 characteristic matrix with the proposed approach.

\begin{example}
Let $U=\{x_{1},x_{2},x_{3},x_{4}\}$,
$\mathscr{C}=\{C_{1},C_{2},C_{3}\}$, where $C_{1}=\{x_{1},x_{4}\}$,
$C_{2}=\{x_{1},x_{2},x_{4}\}$, and $C_{3}=\{x_{3},x_{4}\}$. Then we
have that
$$M_{C_{1}}\cdot M_{C_{1}}^{T}= \left[
\begin{array}{cccc}
1 & 0 & 0 & 1 \\
0 & 0 & 0 & 0 \\
0 & 0 & 0 & 0 \\
1 & 0 & 0 & 1 \\
\end{array}
\right], M_{C_{2}}\cdot M_{C_{2}}^{T}= \left[
\begin{array}{cccc}
1 & 1 & 0 & 1 \\
1 & 1 & 0 & 1 \\
0 & 0 & 0 & 0 \\
1 & 1 & 0 & 1 \\
\end{array}
\right], M_{C_{3}}\cdot M_{C_{3}}^{T}= \left[
\begin{array}{cccc}
0 & 0 & 0 & 0 \\
0 & 0 & 0 & 0 \\
0 & 0 & 1 & 1 \\
0 & 0 & 1 & 1 \\
\end{array}
\right]. $$
Consequently, we obtain that $$M_{\mathscr{C}}\cdot
M_{\mathscr{C}}^{T}=M_{C_{1}}\cdot M_{C_{1}}^{T} \vee M_{C_{2}}\cdot
M_{C_{2}}^{T} \vee M_{C_{3}}\cdot M_{C_{3}}^{T}=\left[
\begin{array}{cccc}
1 & 1 & 0 & 1 \\
1 & 1 & 0 & 1 \\
0 & 0 & 1 & 1 \\
1 & 1 & 1 & 1 \\
\end{array}
\right].$$
\end{example}

Additionally, we discuss that how to get $M_{C}\odot M_{C}^{T}$ with
low time complexity. Actually, we can obtain $M_{C}\odot M_{C}^{T}$
by computing two rows of it, denoted as $(c_{ij})_{n\times n}$.
Concretely, we have that $c_{ik}=c_{jk}$ $(1\leq k\leq n)$ if
$d_{i1}=d_{j1}$. So we only need to compute $c_{ik}$ and $c_{jk}$
$(1\leq k\leq n)$ when $d_{i1}=1$ and $d_{j1}=0$, respectively.

We employ an example to show the process of computing $M_{C}\odot
M_{C}^{T}$ using the proposed approach.

\begin{example}
(Continuation of Example 3.2) Using the proposed approach, we have
that
$$M_{C}\odot M_{C}^{T}=
\left[
\begin{array}{c}
1 \\
0 \\
0 \\
1 \\
\end{array}
\right]\odot\left[
\begin{array}{cccc}
1 & 0 & 0 & 1 \\
\end{array}
\right]=\left[
\begin{array}{cccc}
c_{11} & c_{12} & c_{13} & c_{14} \\
c_{21} & c_{22} & c_{23} & c_{24} \\
c_{31} & c_{32} & c_{33} & c_{34} \\
c_{41} & c_{42} & c_{43} & c_{44} \\
\end{array}
\right]=\left[
\begin{array}{cccc}
1 & 0 & 0 & 1 \\
2 & 1 & 1 & 2 \\
2 & 1 & 1 & 2 \\
1 & 0 & 0 & 1 \\
\end{array}
\right].
$$\end{example}

From the above results, we see that $d_{11}=d_{41}=1$ and
$d_{21}=d_{31}=0$. It is obvious that $c_{1j}=c_{4j}$ and
$c_{2j}=c_{3j}$ $(1\leq j\leq 4)$. To get $M_{C}\odot M_{C}^{T}$, we
can only compute $c_{1j}$ and $c_{2j}$ $(1\leq j\leq 4)$. In this
way, the time complexity of computing $M_{C}\odot M_{C}^{T}$ can be
reduced greatly.

After that, we study the construction of the type-2 characteristic
matrix of the covering based on Example 3.4. Concretely, we obtain
$\{M_{C_{i}}\odot M_{C_{i}}^{T}|C_{i}\in \mathscr{C}\}$ shown in
Table 2. In the sense of Definition 2.8, we have that
$M_{\mathscr{C}}\odot M_{\mathscr{C}}^{T}=\bigwedge_{C_{i}\in
\mathscr{C}} M_{C_{i}}\odot M_{C_{i}}^{T}$. Thus we can get
$M_{\mathscr{C}}\odot M_{\mathscr{C}}^{T}$ by computing
$M_{C_{i}}\odot M_{C_{i}}^{T}$. Furthermore, $\{M_{C_{i}}\odot
M_{C_{i}}^{T}|C_{i}\in \mathscr{C}\}$ are useful for computing the
type-2 characteristic matrix of the dynamic covering, which are
illustrated in Section 4.

\begin{table}[htbp]
\caption{The matrix representations of each $C_{i}\in\mathscr{C}$
and $\mathscr{C}$.} \tabcolsep0.18in
\begin{tabular}{c c c c c c c c}
\hline $U$  &$C_{1}$& $C_{2}$&.&.&. &$C_{m}$& $\mathscr{C}$\\
\hline
$M_{C}\odot M_{C}^{T}$ & $M_{C_{1}}\odot M_{C_{1}}^{T}$& $M_{C_{2}}\odot M_{C_{2}}^{T}$&.&.&. &$M_{C_{m}}\odot M_{C_{m}}^{T}$ & $M_{\mathscr{C}}\odot M_{\mathscr{C}}^{T}$\\
\hline
\end{tabular}
\end{table}

We employ the following example to show the process of computing the
type-2 characteristic matrix with the proposed approach.

\begin{example} (Continuation of Example 3.3)
Similarly, we have that
$$M_{C_{1}}\odot M_{C_{1}}^{T}= \left[
\begin{array}{cccc}
1 & 0 & 0 & 1 \\
2 & 1 & 1 & 2 \\
2 & 1 & 1 & 2 \\
1 & 0 & 0 & 1 \\
\end{array}
\right], M_{C_{2}}\odot M_{C_{2}}^{T}= \left[
\begin{array}{cccc}
1 & 1 & 0 & 1 \\
1 & 1 & 0 & 1 \\
2 & 2 & 1 & 2 \\
1 & 1 & 0 & 1 \\
\end{array}
\right],
M_{C_{3}}\odot M_{C_{3}}^{T}= \left[
\begin{array}{cccc}
1 & 1 & 2 & 2 \\
1 & 1 & 2 & 2 \\
0 & 0 & 1 & 1 \\
0 & 0 & 1 & 1 \\
\end{array}
\right]. $$ Consequently, we obtain that $$M_{\mathscr{C}}\odot
M_{\mathscr{C}}^{T}=M_{C_{1}}\odot M_{C_{1}}^{T} \wedge
M_{C_{2}}\odot M_{C_{2}}^{T} \wedge M_{C_{3}}\odot
M_{C_{3}}^{T}=\left[
\begin{array}{cccc}
1 & 0 & 0 & 1 \\
1 & 1 & 0 & 1 \\
0 & 0 & 1 & 1 \\
0 & 0 & 0 & 1 \\
\end{array}
\right].$$
\end{example}

It is obvious that the time complexity of computing the type-1
(respectively, type-2) characteristic matrix is $m\ast\mathcal
{O}(n)+\mathcal {O}(n^{2})$ if $|U|=n$ and $|\mathscr{C}|=m$. But
the time complexity is $\mathcal {O}(m\ast n^{2})$ by using the
concept of the type-1 (respectively, type-2) characteristic matrix.
Especially, we can compute $M_{C_{i}}\cdot M_{C_{i}}^{T}$ and
$M_{C_{j}}\cdot M_{C_{j}}^{T}$ for $i\neq j$ simultaneously. Thus
the time complexity can be reduced to $\mathcal {O}(n)+\mathcal
{O}(n^{2})$. Therefore, we can get the type-1 and type-2
characteristic matrixes with less time by using the proposed
approach.

In fact, the proposed approaches can be applied to compute
approximations of concepts simultaneously. Concretely, we can
compress the covering approximation space into a small one before
using the second approach.

\section{The construction of characteristic matrixes of the dynamic covering}

In this section, we investigate that how to update the type-1 and
type-2 characteristic matrixes with time. Actually, there exists
five types of dynamic coverings: adding elements into the covering
and deleting some elements of the covering, the immigration and
emigration of object sets, and revising attribute values of some
objects.

\subsection{The characteristic matrixes
of the dynamic covering when varying elements of the covering}

In the following, we introduce the concept of the dynamic covering
approximation space when adding some elements and investigate the
type-1 and type-2 characteristic matrixes of the dynamic covering
when adding some elements into the covering.

\begin{definition}
Let $(U,\mathscr{C})$ be a covering approximation space, where
$U=\{x_{1},x_{2},...,x_{n}\}$ and
$\mathscr{C}=\{C_{1},C_{2},...,\\C_{m}\}$, and
$\mathscr{C}^{\ast}=\{C_{i}^{\ast}|m+1\leq i\leq k\}$. Then
$(U,\mathscr{C})$ is called the original covering approximation
space and $(U,\mathscr{C}^{+})$ is called the AE-covering
approximation space of $(U,\mathscr{C})$, where
$\mathscr{C}^{+}=\mathscr{C}\cup\mathscr{C}^{\ast}$.
\end{definition}

In the sense of Definition 4.1, $\mathscr{C}$ is called the original
covering and $\mathscr{C}^{+}$ is called the AE-covering of the
original covering. Obviously, we can obtain the type-1 and type-2
characteristic matrixes of the AE-covering as the original covering.

We discuss that how to get the type-1 characteristic matrix of
$\mathscr{C}^{+}$ based on that of $\mathscr{C}$. On one hand, it is
obvious that $M_{\mathscr{C}^{+}}\cdot
M_{\mathscr{C}^{+}}^{T}=M_{\mathscr{C}}\cdot M_{\mathscr{C}}^{T}\vee
M_{\mathscr{C}^{\ast}}\cdot M_{\mathscr{C}^{\ast}}^{T}$, so we can
obtain the result by computing $M_{\mathscr{C}^{\ast}}\cdot
M_{\mathscr{C}^{\ast}}^{T}$. But this approach is ineffective for
future computing if some elements of $\mathscr{C}^{+}$ are deleted.
On the other hand, we compute and add $\{M_{C^{\ast}_{i}}\cdot
M_{C^{\ast}_{i}}^{T}|C^{\ast}_{i}\in \mathscr{C}^{\ast}\}$ into
Table 1, and the results are shown in Table 3. In the process of
computing $M_{\mathscr{C}^{+}}\cdot M_{\mathscr{C}^{+}}^{T}$, there
is no need to compute $\{M_{C_{i}}\cdot M_{C_{i}}^{T}|C_{i}\in
\mathscr{C}\}$ by using the results of the original covering.

\begin{table}[htbp]
\caption{The matrix representation of each element of the covering
$\mathscr{C}^{+}$.} \tabcolsep0.19in
\begin{tabular}{c c c c c c c c}
\hline $U$  &$C_{1}$& $C_{2}$&.&.&. &$C_{k}$& $\mathscr{C}^{+}$\\
\hline
$M_{C}\cdot M_{C}^{T}$ & $M_{C_{1}}\cdot M_{C_{1}}^{T}$& $M_{C_{2}}\cdot M_{C_{2}}^{T}$&.&.&. &$M_{C_{k}}\cdot M_{C_{k}}^{T}$ & $M_{\mathscr{C}^{+}}\cdot M_{\mathscr{C}^{+}}^{T}$\\
\hline
\end{tabular}
\end{table}

The following example is employed to show that how to compute the
type-1 characteristic matrix of the AE-covering with the proposed
approach.

\begin{example} (Continuation of Example 3.3)
Let $\mathscr{C}^{+}=\mathscr{C}\cup \{C_{4}\}$, where
$C_{4}=\{x_{2},x_{4}\}$. Obviously, $\mathscr{C}^{+}$ is the
AE-covering of $\mathscr{C}$. To compute $M_{\mathscr{C}^{+}}$, we
only need to compute $M_{C_{4}}\cdot M_{C_{4}}^{T}$ as follows:
$$M_{C_{4}}\cdot M_{C_{4}}^{T}= \left[
\begin{array}{cccc}
0 & 0 & 0 & 0 \\
0 & 1 & 0 & 1 \\
0 & 0 & 0 & 0 \\
0 & 1 & 0 & 1 \\
\end{array}
\right]. $$
Consequently, we obtain that $$M_{\mathscr{C}^{+}}\cdot
M_{\mathscr{C}^{+}}^{T}=M_{C_{1}}\cdot M_{C_{1}}^{T} \vee
M_{C_{2}}\cdot M_{C_{2}}^{T} \vee M_{C_{3}}\cdot M_{C_{3}}^{T}\vee
M_{C_{4}}\cdot M_{C_{4}}^{T}=\left[
\begin{array}{cccc}
1 & 1 & 0 & 1 \\
1 & 1 & 0 & 1 \\
0 & 0 & 1 & 1 \\
1 & 1 & 1 & 1 \\
\end{array}
\right].$$
Furthermore, we can get $$M_{\mathscr{C}^{+}}\cdot
M_{\mathscr{C}^{+}}^{T}=M_{\mathscr{C}}\cdot M_{\mathscr{C}}^{T}
\vee M_{C_{4}}\cdot M_{C_{4}}^{T}=\left[
\begin{array}{cccc}
1 & 1 & 0 & 1 \\
1 & 1 & 0 & 1 \\
0 & 0 & 1 & 1 \\
1 & 1 & 1 & 1 \\
\end{array}
\right].$$
\end{example}

We also investigate that how to construct the type-2 characteristic
matrix of $\mathscr{C}^{+}$ based on that of $\mathscr{C}$. It is
obvious that $M_{\mathscr{C}^{+}}\odot
M_{\mathscr{C}^{+}}^{T}=M_{\mathscr{C}}\odot
M_{\mathscr{C}}^{T}\wedge M_{\mathscr{C}^{\ast}}\odot
M_{\mathscr{C}^{\ast}}^{T}$, so we can obtain the result by
computing $M_{\mathscr{C}^{\ast}}\odot M_{\mathscr{C}^{\ast}}^{T}$.
But this approach is also ineffective for future computing if some
elements of $\mathscr{C}^{+}$ are deleted. Therefore, we focus on
the approach shown in Section 3. Concretely, we compute
$\{M_{C^{\ast}_{i}}\odot M_{C^{\ast}_{i}}^{T}|C^{\ast}_{i}\in
\mathscr{C}^{\ast}\}$ and show the results in Table 4.

\begin{table}[htbp]
\caption{The matrix representation of each element of the covering
$\mathscr{C}^{+}$.} \tabcolsep0.16in
\begin{tabular}{c c c c c c c c}
\hline $U$  &$C_{1}$& $C_{2}$&.&.&. &$C_{k}$& $\mathscr{C}^{+}$\\
\hline
$M_{C}\odot M_{C}^{T}$ & $M_{C_{1}}\odot M_{C_{1}}^{T}$& $M_{C_{2}}\odot M_{C_{2}}^{T}$&.&.&. &$M_{C_{k}}\odot M_{C_{k}}^{T}$ & $M_{\mathscr{C}^{+}}\odot M_{\mathscr{C}^{+}}^{T}$\\
\hline
\end{tabular}
\end{table}

We employ the following example to show the process of computing the
type-2 characteristic matrix of the dynamic covering with the
proposed approach.

\begin{example} (Continuation of Example 4.2) To get $M_{\mathscr{C}^{+}}\odot
M_{\mathscr{C}^{+}}^{T}$, we only need to compute $M_{C_{4}}\odot
M_{C_{4}}^{T}$ as follows:
$$M_{C_{4}}\odot M_{C_{4}}^{T}= \left[
\begin{array}{cccc}
1 & 2 & 1 & 2 \\
0 & 1 & 0 & 1 \\
1 & 2 & 1 & 2 \\
0 & 1 & 0 & 1 \\
\end{array}
\right]. $$
Then we obtain that $$M_{\mathscr{C}^{+}}\odot
M_{\mathscr{C}^{+}}^{T}=M_{C_{1}}\odot M_{C_{1}}^{T} \wedge
M_{C_{2}}\odot M_{C_{2}}^{T} \wedge M_{C_{3}}\odot
M_{C_{3}}^{T}\wedge M_{C_{4}}\odot M_{C_{4}}^{T}=\left[
\begin{array}{cccc}
1 & 0 & 0 & 1 \\
0 & 1 & 0 & 1 \\
0 & 0 & 1 & 1 \\
0 & 0 & 0 & 1 \\
\end{array}
\right].$$ On the other hand, we can get $$M_{\mathscr{C}^{+}}\odot
M_{\mathscr{C}^{+}}^{T}=M_{\mathscr{C}}\odot M_{\mathscr{C}}^{T}
\wedge M_{C_{4}}\odot M_{C_{4}}^{T}=\left[
\begin{array}{cccc}
1 & 0 & 0 & 1 \\
0 & 1 & 0 & 1 \\
0 & 0 & 1 & 1 \\
0 & 0 & 0 & 1 \\
\end{array}
\right].$$
\end{example}

The time complexity of computing the type-1 (respectively, type-2)
characteristic matrix is $(k-m)\ast\mathcal {O}(n)+\mathcal
{O}(n^{2})$ if $|U|=n$ and $|\mathscr{C}^{\ast}|=k$. But the time
complexity is $\mathcal {O}(k\ast n^{2})$ by using the concept of
the type-1 (respectively, type-2) characteristic matrix. Especially,
we can compute $M_{C_{i}}\cdot M_{C_{i}}^{T}$ and $M_{C_{j}}\cdot
M_{C_{j}}^{T}$ for $m\leq i\neq j \leq k$ simultaneously. Thus we
can get the type-1 and type-2 characteristic matrixes with less time
by using the proposed approach.

Subsequently, we introduce the concept of the dynamic covering
approximation space when deleting some elements of the covering and
investigate the type-1 and type-2 characteristic matrixes of the
dynamic covering when deleting some elements of the covering.

\begin{definition}
Let $(U,\mathscr{C})$ be a covering approximation space, where
$U=\{x_{1},x_{2},...,x_{n}\}$,
$\mathscr{C}=\{C_{1},C_{2},...,C_{m}\}$, and
$\mathscr{C}^{\ast}\subseteq \mathscr{C}$. Then
$(U,\mathscr{C}^{-})$ is called the DE-covering approximation space
of $(U,\mathscr{C})$, where
$\mathscr{C}^{-}=\mathscr{C}-\mathscr{C}^{\ast}$.
\end{definition}

In the sense of Definition 4.4, $\mathscr{C}$ is called the original
covering and $\mathscr{C}^{-}$ is called the DE-covering of the
original covering. Furthermore, we can obtain the type-1 and type-2
characteristic matrixes of the DE-covering as the original covering.

We discuss that how to get the type-1 and type-2 characteristic
matrixes of $\mathscr{C}^{-}$ based on those of $\mathscr{C}$. There
are two steps to compute the type-1 and type-2 characteristic
matrixes of $\mathscr{C}^{-}$. To express clearly, we assume that
only $C_{N} \in \mathscr{C}$ is deleted. First, we get Tables 5 and
6 by deleting $M_{C_{N}}\cdot M_{C_{N}}^{T}$ and $M_{C_{N}}\odot
M_{C_{N}}^{T}$ in Tables 2 and 3, respectively. Second, we obtain
that $M_{\mathscr{C}^{-}}\cdot
M_{\mathscr{C}^{-}}^{T}=\bigvee_{C_{i}\in \mathscr{C}^{-}}
M_{C_{i}}\cdot M_{C_{i}}^{T}$ and $M_{\mathscr{C}^{-}}\odot
M_{\mathscr{C}^{-}}^{T}=\bigwedge_{C_{i}\in \mathscr{C}^{-}}
M_{C_{i}}\odot M_{C_{i}}^{T}$. Clearly, there is no need to compute
$\{M_{C_{i}}\cdot M_{C_{i}}^{T}|C_{i}\in \mathscr{C}-\{C_{N}\}\}$
and $\{M_{C_{i}}\odot M_{C_{i}}^{T}|C_{i}\in
\mathscr{C}-\{C_{N}\}\}$. Similarly, we can get the type-1 and
type-2 characteristic matrixes of the DE-covering when a set of
elements of the original covering is deleted simultaneously.

\begin{table}[htbp]
\caption{The matrix representation of each element of the covering
$\mathscr{C}^{-}$.} \tabcolsep0.05in
\begin{tabular}{c c c c c c c c c c c c c c}
\hline $U$  &$C_{1}$& $C_{2}$&.&.&.&$C_{N-1}$&$C_{N+1}$&.&.&.&$C_{m}$& $\mathscr{C}^{-}$\\
\hline
$M_{C}\cdot M_{C}^{T}$ & $M_{C_{1}}\cdot M_{C_{1}}^{T}$& $M_{C_{2}}\cdot M_{C_{2}}^{T}$&.&.&.& $M_{C_{N-1}}\cdot M_{C_{N-1}}^{T}$ & $M_{C_{N+1}}\cdot M_{C_{N+1}}^{T}$ &.&.&.&$M_{C_{m}}\cdot M_{C_{m}}^{T}$ & $M_{\mathscr{C}^{-}}\cdot M_{\mathscr{C}^{-}}^{T}$\\
\hline
\end{tabular}
\end{table}

\begin{table}[htbp]
\caption{The matrix representation of each element of the covering
$\mathscr{C}^{-}$.} \tabcolsep0.035in
\begin{tabular}{c c c c c c c c c c c c c c}
\hline $U$  &$C_{1}$& $C_{2}$&.&.&.&$C_{N-1}$&$C_{N+1}$&.&.&.&$C_{m}$& $\mathscr{C}^{-}$\\
\hline
$M_{C}\odot M_{C}^{T}$ & $M_{C_{1}}\odot M_{C_{1}}^{T}$& $M_{C_{2}}\odot M_{C_{2}}^{T}$&.&.&.& $M_{C_{N-1}}\odot M_{C_{N-1}}^{T}$ & $M_{C_{N+1}}\odot M_{C_{N+1}}^{T}$ &.&.&.&$M_{C_{m}}\odot M_{C_{m}}^{T}$ & $M_{\mathscr{C}^{-}}\odot M_{\mathscr{C}^{-}}^{T}$\\
\hline
\end{tabular}
\end{table}

The following example is employed to show that how to compute the
type-1 and type-2 characteristic matrixes of the DE-covering with
the proposed approach.

\begin{example} (Continuation of Example 3.3)
Let $\mathscr{C}^{-}=\mathscr{C}- \{C_{3}\}$, and $\mathscr{C}^{-}$
is the DE-covering of $\mathscr{C}$. Then we have that
\begin{eqnarray*}M_{\mathscr{C}^{-}}\cdot
M_{\mathscr{C}^{-}}^{T}&=&M_{C_{1}}\cdot M_{C_{1}}^{T} \vee
M_{C_{2}}\cdot M_{C_{2}}^{T} =\left[
\begin{array}{cccc}
1 & 1 & 0 & 1 \\
1 & 1 & 0 & 1 \\
0 & 0 & 0 & 0 \\
1 & 1 & 0 & 1 \\
\end{array}
\right];\\
M_{\mathscr{C}^{-}}\odot M_{\mathscr{C}^{-}}^{T}&=&M_{C_{1}}\odot
M_{C_{1}}^{T} \wedge M_{C_{2}}\odot M_{C_{2}}^{T} =\left[
\begin{array}{cccc}
1 & 0 & 0 & 1 \\
1 & 1 & 0 & 1 \\
2 & 1 & 1 & 2 \\
1 & 0 & 0 & 1 \\
\end{array}
\right].
\end{eqnarray*}
\end{example}

In the process of computing $M_{\mathscr{C}^{-}}\cdot
M_{\mathscr{C}^{-}}^{T}$ and $M_{\mathscr{C}^{-}}\odot
M_{\mathscr{C}^{-}}^{T}$, there is no need to compute
$M_{C_{1}}\cdot M_{C_{1}}^{T} $, $M_{C_{2}}\cdot M_{C_{2}}^{T}$,
$M_{C_{1}}\odot M_{C_{1}}^{T} $ and $M_{C_{2}}\odot M_{C_{2}}^{T}$.
Thereby, the computing of the type-1 and type-2 characteristic
matrixes of the DE-covering can be reduced greatly.

The time complexity of computing the type-1 (respectively, type-2)
characteristic matrix is $\mathcal {O}(n^{2})$ if $|U|=n$ and
$|\mathscr{C}^{-}|=k$. But the time complexity is $\mathcal
{O}(k\ast n^{2})$ by using the concept of the type-1 (respectively,
type-2) characteristic matrix.

\subsection{The characteristic matrixes
of the dynamic covering when adding some new objects and deleting
some objects}

In the following, we introduce the concept of the dynamic covering
approximation space when adding some objects and investigate the
type-1 and type-2 characteristic matrixes of the dynamic covering
when adding some objects.

\begin{definition}
Let $(U,\mathscr{C})$ be a covering approximation space, where
$U=\{x_{1},x_{2},...,x_{n}\}$ and
$\mathscr{C}=\{C_{1},C_{2},...,\\C_{m}\}$, $U^{\ast}=\{x_{i}|n+1\leq
i\leq n+t\}$, $U^{+}=U\cup U^{\ast},$
$\mathscr{C}^{+}=\{C^{+}_{1},C^{+}_{2},...,C^{+}_{m}\}$, $C^{+}_{i}$
is updated by adding some new objects into $C_{i}$. Then
$(U^{+},\mathscr{C}^{+})$ is called the AO-covering approximation
space of $(U,\mathscr{C})$.
\end{definition}

In the sense of Definition 4.6, $\mathscr{C}$ is called the original
covering and $\mathscr{C}^{+}$ is called the updated covering of the
original covering. Furthermore, we can obtain the type-1 and type-2
characteristic matrixes of the updated covering as the original
covering.

We discuss that how to get the type-1 characteristic matrix of
$\mathscr{C}^{+}$ based on that of $\mathscr{C}$. Concretely, we
study the relationship between
$\Gamma(\mathscr{C})=(b_{ij})_{(n+t)\times (n+t)}$ and
$\Gamma(\mathscr{C}^{+})=(c_{ij})_{(n+t)\times (n+t)}$. It is
obvious that $\Gamma(\mathscr{C}^{+})$ is symmetric and
$b_{ij}=c_{ij}$ $(1\leq i,j\leq n)$. Thus we only need to compute
$c_{ij}$ $(n+1\leq i\leq n+t,1\leq j\leq n+t)$, denoted as
$\triangle\Gamma(\mathscr{C})$. Therefore, the time complexity of
computing the type-1 characteristic matrix of $\mathscr{C}^{+}$ can
be reduced greatly.

\begin{eqnarray*}
\Gamma(\mathscr{C})=M_{\mathscr{C}}\cdot
M_{\mathscr{C}}^{T}&=&\left[
  \begin{array}{cccccc}
    a_{11} & a_{12} & . & . & . & a_{1m} \\
    a_{21} & a_{22} & . & . & . & a_{2m} \\
    . & . & . & . & . & . \\
    . & . & . & . & . & . \\
    . & . & . & . & . & . \\
    a_{n1} & a_{n2} & . & . & . & a_{nm} \\
  \end{array}
\right] \cdot \left[
  \begin{array}{cccccc}
    a_{11} & a_{12} & . & . & . & a_{1m} \\
    a_{21} & a_{22} & . & . & . & a_{2m} \\
    . & . & . & . & . & . \\
    . & . & . & . & . & . \\
    . & . & . & . & . & . \\
    a_{n1} & a_{n2} & . & . & . & a_{nm} \\
  \end{array}
\right]^{T}\\
&=&\left[
  \begin{array}{cccccc}
    b_{11} & b_{12} & . & . & . & b_{1n} \\
    b_{21} & b_{22} & . & . & . & b_{2n} \\
    . & . & . & . & . & . \\
    . & . & . & . & . & . \\
    . & . & . & . & . & . \\
    b_{n1} & b_{n2} & . & . & . & b_{nn} \\
  \end{array}
\right];
\\
\Gamma(\mathscr{C}^{+})=M_{\mathscr{C}^{+}}\cdot
M_{\mathscr{C}^{+}}^{T}&=&\left[
  \begin{array}{cccccc}
    a_{11} & a_{12} & . & . & . & a_{1m} \\
    a_{21} & a_{22} & . & . & . & a_{2m} \\
    . & . & . & . & . & . \\
    . & . & . & . & . & . \\
    . & . & . & . & . & . \\
    a_{n1} & a_{n2} & . & . & . & a_{nm} \\
    a_{(n+1)1} & a_{(n+1)2} & . & . & . & a_{(n+1)m} \\
    . & . & . & . & . & . \\
    . & . & . & . & . & . \\
    . & . & . & . & . & . \\
    a_{(n+t)1} & a_{(n+t)2} & . & . & . & a_{(n+t)m} \\
  \end{array}
\right]\cdot \left[
  \begin{array}{cccccc}
    a_{11} & a_{12} & . & . & . & a_{1m} \\
    a_{21} & a_{22} & . & . & . & a_{2m} \\
    . & . & . & . & . & . \\
    . & . & . & . & . & . \\
    . & . & . & . & . & . \\
    a_{n1} & a_{n2} & . & . & . & a_{nm} \\
    a_{(n+1)1} & a_{(n+1)2} & . & . & . & a_{(n+1)m} \\
    . & . & . & . & . & . \\
    . & . & . & . & . & . \\
    . & . & . & . & . & . \\
    a_{(n+t)1} & a_{(n+t)2} & . & . & . & a_{(n+t)m} \\
  \end{array}
\right]^{T} \\ &=&\left[
  \begin{array}{ccccccccccc}
    c_{11} & c_{12} & . & . & . & c_{1n}& c_{1(n+1)}&.&.&.& c_{1(n+t)} \\
    c_{21} & c_{22} & . & . & . & c_{2n}& c_{2(n+1)}&.&.&.& c_{2(n+t)} \\
    . & . & . & . & . & . & . & . & . & . & .\\
    . & . & . & . & . & . & . & . & . & . & .\\
    . & . & . & . & . & . & . & . & . & . & .\\
    c_{n1} & c_{n2} & . & . & .& c_{nn}& c_{n(n+1)}&.&.&. & c_{n(n+t)} \\
    c_{(n+1)1} & c_{(n+1)2} & . & . & . & c_{(n+1)n}& c_{(n+1)(n+1)}&.&.&.& c_{(n+1)(n+t)} \\
    . & . & . & . & . & . & . & . & . & . & .\\
    . & . & . & . & . & . & . & . & . & . & .\\
    . & . & . & . & . & . & . & . & . & . & .\\
    c_{(n+t)1} & c_{(n+t)2} & . & . & . & c_{(n+t)n}& c_{(n+t)(n+1)}&.&.&.& c_{(n+t)(n+t)} \\
  \end{array}
\right];
\\
\triangle\Gamma(\mathscr{C})&=&\left[
  \begin{array}{ccccccccccc}
    c_{(n+1)1} & c_{(n+1)2} & . & . & . & c_{(n+1)n}& c_{(n+1)(n+1)}&.&.&.& c_{(n+1)(n+t)} \\
    c_{(n+2)1} & c_{(n+2)2} & . & . & . & c_{(n+2)n}& c_{(n+2)(n+1)}&.&.&.& c_{(n+2)(n+t)} \\
    . & . & . & . & . & . & . & . & . & . & .\\
    . & . & . & . & . & . & . & . & . & . & .\\
    . & . & . & . & . & . & . & . & . & . & .\\
    c_{(n+t)1} & c_{(n+t)2} & . & . & . & c_{(n+t)n}& c_{(n+t)(n+1)}&.&.&.& c_{(n+t)(n+t)} \\
  \end{array}
\right]
\end{eqnarray*}
\begin{eqnarray*}
 &=& \left[
  \begin{array}{cccccc}
    a_{(n+1)1} & a_{(n+1)2} & . & . & . & a_{(n+1)m} \\
    a_{(n+2)1} & a_{(n+2)2} & . & . & . & a_{(n+2)m} \\
    . & . & . & . & . & . \\
    . & . & . & . & . & . \\
    . & . & . & . & . & . \\
    a_{(n+t)1} & a_{(n+t)2} & . & . & . & a_{(n+t)m} \\
  \end{array}
\right]\cdot \left[
  \begin{array}{cccccc}
    a_{11} & a_{12} & . & . & . & a_{1m} \\
    a_{21} & a_{22} & . & . & . & a_{2m} \\
    . & . & . & . & . & . \\
    . & . & . & . & . & . \\
    . & . & . & . & . & . \\
    a_{n1} & a_{n2} & . & . & . & a_{nm} \\
    a_{(n+1)1} & a_{(n+1)2} & . & . & . & a_{(n+1)m} \\
    . & . & . & . & . & . \\
    . & . & . & . & . & . \\
    . & . & . & . & . & . \\
    a_{(n+t)1} & a_{(n+t)2} & . & . & . & a_{(n+t)m} \\
  \end{array}
\right]^{T}.
\end{eqnarray*}

The following example is employed to show that how to compute the
type-1 characteristic matrix of the dynamic covering with the
proposed approach.

\begin{example} (Continuation of Example 3.3)
Let $U^{+}=U\cup\{x_{5},x_{6}\}$,
$\mathscr{C}^{+}=\{C^{+}_{1},C^{+}_{2},C^{+}_{3}\}$, where
$C^{+}_{1}=\{x_{1},x_{4},x_{5}\}$,
$C^{+}_{2}=\{x_{1},x_{2},x_{4},x_{5},x_{6}\}$, and
$C^{+}_{3}=\{x_{3},x_{4},x_{6}\}$, and $\mathscr{C}^{-}$ is the
AO-covering of $\mathscr{C}$. Then we obtain that
$$\triangle\Gamma(\mathscr{C})=\left[
\begin{array}{ccc}
1 & 1 & 0 \\
0 & 1 & 1 \\
\end{array}
\right]\cdot \left[
\begin{array}{cccccc}
1 & 0 & 0 & 1 &1 &0\\
1 & 1 & 0 & 1 &1 &1\\
0 & 0 & 1 & 1 &0 &1\\
\end{array}
\right]=\left[
\begin{array}{cccccc}
1 & 1 & 0 & 1 &1 &1\\
1 & 1 & 0 & 1 &1 &1\\
\end{array}
\right].$$
\end{example}
Therefore, we obtain $\Gamma(\mathscr{C}^{+})=\left[
\begin{array}{cccccc}
1 & 1 & 0 & 1 &1 & 1\\
1 & 1 & 0 & 1 &1 & 1\\
0 & 0 & 1 & 1 &0 & 0\\
1 & 1 & 1 & 1 &1 & 1\\
1 & 1 & 0 & 1 &1 & 1\\
1 & 1 & 0 & 1 &1 & 1\\
\end{array}
\right].$

Then, we discuss that how to get the type-2 characteristic matrix of
$\mathscr{C}^{+}$ based on that of $\mathscr{C}$. We study the
relationship between $\prod(\mathscr{C})=(b_{ij})_{n+t)\times n+t)}$
and $\prod(\mathscr{C}^{+})=(c_{ij})_{n+t)\times n+t)}$. It is
obvious that $b_{ij}=c_{ij}$ $(1\leq i,j\leq n)$. Thus we only need
to compute $\bigtriangleup_{1}\prod= (c_{ij})_{(n+1\leq i\leq
n+t,1\leq j\leq n+t)}$ and $\bigtriangleup_{2}\prod=
(c_{ij})_{(1\leq i\leq n,n+1\leq j\leq n+t)}$. Therefore, the time
complexity of computing the type-2 characteristic matrix of
$\mathscr{C}^{+}$ can be reduced greatly.
\begin{eqnarray*}
\prod(\mathscr{C})=M_{\mathscr{C}}\odot M_{\mathscr{C}}^{T}&=&\left[
  \begin{array}{cccccc}
    a_{11} & a_{12} & . & . & . & a_{1m} \\
    a_{21} & a_{22} & . & . & . & a_{2m} \\
    . & . & . & . & . & . \\
    . & . & . & . & . & . \\
    . & . & . & . & . & . \\
    a_{n1} & a_{n2} & . & . & . & a_{nm} \\
  \end{array}
\right] \odot \left[
  \begin{array}{cccccc}
    a_{11} & a_{12} & . & . & . & a_{1m} \\
    a_{21} & a_{22} & . & . & . & a_{2m} \\
    . & . & . & . & . & . \\
    . & . & . & . & . & . \\
    . & . & . & . & . & . \\
    a_{n1} & a_{n2} & . & . & . & a_{nm} \\
  \end{array}
\right]^{T}
\\
&=&\left[
  \begin{array}{cccccc}
    b_{11} & b_{12} & . & . & . & b_{1n} \\
    b_{21} & b_{22} & . & . & . & b_{2n} \\
    . & . & . & . & . & . \\
    . & . & . & . & . & . \\
    . & . & . & . & . & . \\
    b_{n1} & b_{n2} & . & . & . & b_{nn} \\
  \end{array}
\right];
\end{eqnarray*}
\begin{eqnarray*}
\prod(\mathscr{C}^{+})=M_{\mathscr{C}^{+}}\odot
M_{\mathscr{C}^{+}}^{T}&=&\left[
  \begin{array}{cccccc}
    a_{11} & a_{12} & . & . & . & a_{1m} \\
    a_{21} & a_{22} & . & . & . & a_{2m} \\
    . & . & . & . & . & . \\
    . & . & . & . & . & . \\
    . & . & . & . & . & . \\
    a_{n1} & a_{n2} & . & . & . & a_{nm} \\
    a_{(n+1)1} & a_{(n+1)2} & . & . & . & a_{(n+1)m} \\
    . & . & . & . & . & . \\
    . & . & . & . & . & . \\
    . & . & . & . & . & . \\
    a_{(n+t)1} & a_{(n+t)2} & . & . & . & a_{(n+t)m} \\
  \end{array}
\right]\odot \left[
  \begin{array}{cccccc}
    a_{11} & a_{12} & . & . & . & a_{1m} \\
    a_{21} & a_{22} & . & . & . & a_{2m} \\
    . & . & . & . & . & . \\
    . & . & . & . & . & . \\
    . & . & . & . & . & . \\
    a_{n1} & a_{n2} & . & . & . & a_{nm} \\
    a_{(n+1)1} & a_{(n+1)2} & . & . & . & a_{(n+1)m} \\
    . & . & . & . & . & . \\
    . & . & . & . & . & . \\
    . & . & . & . & . & . \\
    a_{(n+t)1} & a_{(n+t)2} & . & . & . & a_{(n+t)m} \\
  \end{array}
\right]^{T}
\\
&=&\left[
  \begin{array}{ccccccccccc}
    c_{11} & c_{12} & . & . & . & c_{1n}& c_{1(n+1)}&.&.&.& c_{1(n+t)} \\
    c_{21} & c_{22} & . & . & . & c_{2n}& c_{2(n+1)}&.&.&.& c_{2(n+t)} \\
    . & . & . & . & . & . & . & . & . & . & .\\
    . & . & . & . & . & . & . & . & . & . & .\\
    . & . & . & . & . & . & . & . & . & . & .\\
    c_{n1} & c_{n2} & . & . & .& c_{nn}& c_{n(n+1)}&.&.&. & c_{n(n+t)} \\
    c_{(n+1)1} & c_{(n+1)2} & . & . & . & c_{(n+1)n}& c_{(n+1)(n+1)}&.&.&.& c_{(n+1)(n+t)} \\
    . & . & . & . & . & . & . & . & . & . & .\\
    . & . & . & . & . & . & . & . & . & . & .\\
    . & . & . & . & . & . & . & . & . & . & .\\
    c_{(n+t)1} & c_{(n+t)2} & . & . & . & c_{(n+t)n}& c_{(n+t)(n+1)}&.&.&.& c_{(n+t)(n+t)} \\
  \end{array}
\right];
\\
\triangle_{1}\prod(\mathscr{C})&=&\left[
  \begin{array}{ccccccccccc}
    c_{(n+1)1} & c_{(n+1)2} & . & . & . & c_{(n+1)n}& c_{(n+1)(n+1)}&.&.&.& c_{(n+1)(n+t)} \\
    c_{(n+2)1} & c_{(n+2)2} & . & . & . & c_{(n+2)n}& c_{(n+2)(n+1)}&.&.&.& c_{(n+2)(n+t)} \\
    . & . & . & . & . & . & . & . & . & . & .\\
    . & . & . & . & . & . & . & . & . & . & .\\
    . & . & . & . & . & . & . & . & . & . & .\\
    c_{(n+t)1} & c_{(n+t)2} & . & . & . & c_{(n+t)n}& c_{(n+t)(n+1)}&.&.&.& c_{(n+t)(n+t)} \\
  \end{array}
\right] \\ &=& \left[
  \begin{array}{cccccc}
    a_{(n+1)1} & a_{(n+1)2} & . & . & . & a_{(n+1)m} \\
    a_{(n+2)1} & a_{(n+2)2} & . & . & . & a_{(n+2)m} \\
    . & . & . & . & . & . \\
    . & . & . & . & . & . \\
    . & . & . & . & . & . \\
    a_{(n+t)1} & a_{(n+t)2} & . & . & . & a_{(n+t)m} \\
  \end{array}
\right]\odot \left[
  \begin{array}{cccccc}
    a_{11} & a_{12} & . & . & . & a_{1m} \\
    a_{21} & a_{22} & . & . & . & a_{2m} \\
    . & . & . & . & . & . \\
    . & . & . & . & . & . \\
    . & . & . & . & . & . \\
    a_{n1} & a_{n2} & . & . & . & a_{nm} \\
    a_{(n+1)1} & a_{(n+1)2} & . & . & . & a_{(n+1)m} \\
    . & . & . & . & . & . \\
    . & . & . & . & . & . \\
    . & . & . & . & . & . \\
    a_{(n+t)1} & a_{(n+t)2} & . & . & . & a_{(n+t)m} \\
  \end{array}
\right]^{T};
\\
 \triangle_{2}\prod(\mathscr{C})&=&\left[
  \begin{array}{ccccccccccc}
    c_{1(n+1)} & c_{2(n+2)} & . & . & . & c_{n(n+t)}\\
    c_{2(n+1)} & c_{2(n+2)} & . & . & . & c_{2(n+t)}\\
    . & . & . & . & . \\
    . & . & . & . & . \\
    . & . & . & . & . \\
    c_{1(n+1)} & c_{2(n+2)} & . & . & . & c_{n(n+t)}\\
  \end{array}
\right] \\
&=& \left[
  \begin{array}{cccccc}
    a_{11} & a_{12} & . & . & . & a_{1m} \\
    a_{21} & a_{22} & . & . & . & a_{2m} \\
    . & . & . & . & . & . \\
    . & . & . & . & . & . \\
    . & . & . & . & . & . \\
    a_{n1} & a_{n2} & . & . & . & a_{nm} \\
  \end{array}
\right]\odot \left[
  \begin{array}{cccccc}
    a_{(n+1)1} & a_{(n+1)2} & . & . & . & a_{(n+1)m} \\
    a_{(n+2)1} & a_{(n+2)2} & . & . & . & a_{(n+2)m} \\
    . & . & . & . & . & . \\
    . & . & . & . & . & . \\
    . & . & . & . & . & . \\
    a_{(n+t)1} & a_{(n+t)2} & . & . & . & a_{(n+t)m} \\
  \end{array}
\right]^{T}.
\end{eqnarray*}

The following example is employed to show that how to compute the
type-1 characteristic matrix of the dynamic covering with the
proposed approach.

\begin{example} (Continuation of Example 4.7)
By Definition 4.6, $\mathscr{C}^{-}$ is the AO-covering of
$\mathscr{C}$. Thus we obtain that
\begin{eqnarray*}
\triangle_{1}\prod(\mathscr{C})&=&\left[
\begin{array}{ccc}
1 & 1 & 0 \\
0 & 1 & 1 \\
\end{array}
\right]\odot \left[
\begin{array}{cccccc}
1 & 0 & 0 & 1 &1 &0\\
1 & 1 & 0 & 1 &1 &1\\
0 & 0 & 1 & 1 &0 &1\\
\end{array}
\right]=\left[
\begin{array}{cccccc}
1 & 0 & 0 & 1 &1 &0\\
0 & 0 & 0 & 1 &0 &1\\
\end{array}
\right];
\\
\triangle_{2}\prod(\mathscr{C})&=&\left[
\begin{array}{ccc}
1 & 1 & 0 \\
0 & 1 & 0 \\
0 & 0 & 1 \\
1 & 1 & 1 \\
\end{array}
\right]\odot \left[
\begin{array}{cc}
1 & 0 \\
1 & 1 \\
0 & 1 \\
\end{array}
\right]=\left[
\begin{array}{cccccc}
1 & 0 \\
1 & 1 \\
0 & 1 \\
0 & 0 \\
\end{array}
\right].\end{eqnarray*}
\end{example}

Therefore, we obtain $\prod(\mathscr{C}^{+})=\left[
\begin{array}{cccccc}
1 & 0 & 0 & 1 & 1 & 0\\
1 & 1 & 0 & 1 & 1 & 1\\
0 & 0 & 1 & 1 & 0 & 1\\
0 & 0 & 0 & 1 & 0 & 0\\
1 & 0 & 0 & 1 & 1 & 0\\
0 & 0 & 0 & 1 & 0 & 1\\
\end{array}
\right].$

The time complexity of computing the type-1 (respectively, type-2)
characteristic matrix is $\mathcal {O}((n+t)^{2})+\mathcal {O}(m\ast
t\ast(n+t))$ if $|U^{\ast}|=n+t$ and $|\mathscr{C}|=m$. But the time
complexity is $\mathcal {O}(m\ast (n+t)^{2})$ by using the concept
of the type-1 (respectively, type-2) characteristic matrix.
Therefore, we can get the type-1 and type-2 characteristic matrixes
with less time by using the proposed approach.

Subsequently, we introduce the concept of the dynamic covering
approximation space when deleting some objects and investigate the
type-1 and type-2 characteristic matrixes of the dynamic covering
when deleting some objects.

\begin{definition}
Let $(U,\mathscr{C})$ be a covering approximation space, where
$U=\{x_{1},x_{2},...,x_{n}\}$ and
$\mathscr{C}=\{C_{1},C_{2},...,\\C_{m}\}$, $U^{\ast}\subseteq U$,
$U^{-}=U-U^{\ast},$
$\mathscr{C}^{-}=\{C^{-}_{1},C^{-}_{2},...,C^{-}_{m}\}$, and
$C^{-}_{i}$ is updated by deleting some objects of $C_{i}$ which
belong to $U^{\ast}$. Then $(U^{-},\mathscr{C}^{-})$ is called the
DO-covering approximation space of $(U,\mathscr{C})$.
\end{definition}

In the sense of Definition 4.9, $\mathscr{C}$ is called the original
covering and $\mathscr{C}^{-}$ is called the DO-covering of the
original covering. Clearly, we can obtain the type-1 and type-2
characteristic matrixes of the DO-covering as the original covering.

We investigate that how to obtain the type-1 characteristic matrix
$\Gamma(\mathscr{C}^{-})$ and the type-2 characteristic matrix
$\prod(\mathscr{C}^{-})$ of the DO-covering. Suppose that we have
obtained $\Gamma(\mathscr{C})=(b_{ij})_{n\times n}$ and
$\prod(\mathscr{C})=(c_{ij})_{n\times n}$ shown in Section 3. If we
delete the objects $\{x_{i_{k}}|1\leq k \leq N\}$, actually, it is
easy to get $\Gamma(\mathscr{C}^{-})$ and $\prod(\mathscr{C}^{-})$
based on $\Gamma(\mathscr{C})$ and $\prod(\mathscr{C})$,
respectively. Concretely, we can obtain $\Gamma(\mathscr{C}^{-})$ by
deleting the elements $\{b_{i_{k}j}\}$, $\{b_{ji_{k}}\}$,
$\{c_{i_{k}j}\}$ and $\{c_{ji_{k}}\}$ $(1\leq j \leq n, 1\leq k\leq
N)$ in $\Gamma(\mathscr{C})$ and $\prod(\mathscr{C})$, respectively.

We employ an example to show that how to get the type-1 and type-2
characteristic matrixes of the DO-covering when deleting some
objects.

\begin{example} (Continuation of Example 3.3)
Suppose that we delete the object $x_{4}$, it is obvious that
$U^{-}=\{x_{1},x_{2},x_{3}\}$,
$\mathscr{C}^{-}=\{C^{-}_{1},C^{-}_{2},C^{-}_{4}\}$, where
$C^{-}_{1}=\{x_{1}\}$, $C^{-}_{2}=\{x_{1},x_{2}\}$ and
$C^{-}_{3}=\{x_{3}\}$. On one hand, we can delete the element
$b_{41},b_{42},b_{43},b_{44},b_{14},b_{24},b_{34}$ in the matrix
$\Gamma(\mathscr{C})$ shown in Example 3.3 and
$c_{41},c_{42},c_{43},c_{44},c_{14},c_{24},c_{34}$ in the matrix
$\prod(\mathscr{C})$ shown in Example 3.5. Then we have that
\begin{eqnarray*}\Gamma(\mathscr{C}^{-})&=&\left[
\begin{array}{ccc}
1 & 1 & 0 \\
1 & 1 & 0 \\
0 & 0 & 1 \\
\end{array}
\right];\\
\prod(\mathscr{C}^{-})&=&\left[
\begin{array}{ccc}
1 & 0 & 0 \\
1 & 1 & 0 \\
0 & 0 & 1 \\
\end{array}
\right].\end{eqnarray*} On the other hand, we can get the above
results as $\Gamma(\mathscr{C})$ and $\prod(\mathscr{C})$ shown in
Examples 3.3 and 3.5, respectively:
\begin{eqnarray*}\Gamma(\mathscr{C}^{-})&=&M_{\mathscr{C}^{-}}\cdot
M_{\mathscr{C}^{-}}^{T}=\left[
\begin{array}{ccc}
1 & 1 & 0 \\
1 & 1 & 0 \\
0 & 0 & 1 \\
\end{array}
\right];\\
\prod(\mathscr{C}^{-})&=&M_{\mathscr{C}^{-}}\odot
M_{\mathscr{C}^{-}}^{T}=\left[
\begin{array}{ccc}
1 & 0 & 0 \\
1 & 1 & 0 \\
0 & 0 & 1 \\
\end{array}
\right].\end{eqnarray*}
\end{example}

The time complexity of computing the type-1 (respectively, type-2)
characteristic matrix is $\mathcal {O}((n-t)^{2})$ if
$|U^{\ast}|=n-t$ and $|\mathscr{C}|=m$. But the time complexity is
$\mathcal {O}(m\ast (n-t)^{2})$ by using the concept of the type-1
(respectively, type-2) characteristic matrix.

\subsection{The characteristic matrixes
of the dynamic covering when changing the attribute values of
objects}

In this subsection, we introduce the concept of the dynamic covering
approximation space when changing the attribute values of some
objects and investigate the type-1 and type-2 characteristic
matrixes of the dynamic covering when there are changes of the
attribute values for some objects.

\begin{definition}
Let $(U,\mathscr{C})$ be a covering approximation space, where
$U=\{x_{1},x_{2},...,x_{n}\}$ and
$\mathscr{C}=\{C_{1},C_{2},...,C_{m}\}$. If we cancle some objects
$x$ from $C \in\mathscr{C}$ and add them into other element of $
\mathscr{C}$, and
$\mathscr{C}^{\ast}=\{C^{\ast}_{1},C^{\ast}_{2},...,C^{\ast}_{m}\}$,
where $C^{\ast}_{i}$ is the updated version of $C_{i}$ by deleting
or adding objects. Then $(U,\mathscr{C}^{\ast})$ is called the
CA-covering approximation space of $(U,\mathscr{C})$.
\end{definition}

Generally speaking, if some attribute values are revised  with time
in the original approximation space $(U,\mathscr{C})$, then we will
get the CA-covering approximation space $(U,\mathscr{C}^{\ast})$.
Maybe we get $|\mathscr{C}|<|\mathscr{C}^{\ast}|$ if the change of
attribute values of objects results in that the objects do not
belong to the existing element of $\mathscr{C}$, and we will discuss
the the above situation at the end of this subsection.

We focus on investigating the situation that
$|\mathscr{C}|=|\mathscr{C}^{\ast}|$. Concretely, we only discuss
that an object is deleted in an element of $\mathscr{C}$ and added
into another element of $\mathscr{C}$. Suppose that we have got
$\Gamma(\mathscr{C})$, and the attribute value of $x_{k}\in U$ is
revised and it is deleted in $C_{I}\in \mathscr{C}$ and added into
$C_{J}\in\mathscr{C}$. We study the relationship between
$\Gamma(\mathscr{C})$ and $\Gamma(\mathscr{C}^{\ast})$ by showing
them as follows. Concretely, we see that $b_{ij}=c_{ij}$ for $i\neq
k$ and $ j\neq k$. Since $c_{ki}=c_{ik}$ in
$\Gamma(\mathscr{C}^{\ast})$, we only need to compute $c_{kj}$ for
$1\leq j\leq n$.

\begin{eqnarray*}
\Gamma(\mathscr{C})=M_{\mathscr{C}}\cdot
M^{T}_{\mathscr{C}}&=&\left[
  \begin{array}{cccccccccccccc}
    a_{11} & a_{12} & . & .& . & a_{1I} & . & . & . & a_{1J} & . & . & . & a_{1m} \\
    a_{21} & a_{22} & . & .& . & a_{2I} & . & . & . & a_{2J} & . & . & . & a_{2m} \\
    . & . & . & . & . & . & . & . & . & . & . & .& . & . \\
    . & . & . & . & . & . & . & . & . & . & . & . & .& . \\
    . & . & . & . & . & . & . & . & . & . & . & . & .& . \\
    a_{k1} & a_{k2} & . & . & . & 1 & . & . & . & 0 & . & . & . & a_{km} \\
    . & . & . & . & . & . & . & . & . & . & . & . & . & . \\
    . & . & . & . & . & . & . & . & . & . & . & . & . & . \\
    . & . & . & . & . & . & .&  . & . & . & . & . & . & . \\
    a_{n1} & a_{n2} & . & . & . &a_{kI}& . & . & . & a_{kJ} & . & . & . & a_{nm} \\
  \end{array}
\right]
\\
&\cdot& \left[
  \begin{array}{cccccccccccccc}
    a_{11} & a_{12} & . & .& . & a_{1I} & . & . & . & a_{1J} & . & . & . & a_{1m} \\
    a_{21} & a_{22} & . & .& . & a_{2I} & . & . & . & a_{2J} & . & . & . & a_{2m} \\
    . & . & . & . & . & . & . & . & . & . & . & .& . & . \\
    . & . & . & . & . & . & . & . & . & . & . & . & .& . \\
    . & . & . & . & . & . & . & . & . & . & . & . & .& . \\
    a_{k1} & a_{k2} & . & . & .& 1 & . & . & . & 0 & . & . & . & a_{km} \\
    . & . & . & . & . & . & . & . & . & . & . & . & . & . \\
    . & . & . & . & . & . & . & . & . & . & . & . & . & . \\
    . & . & . & . & . & . & .&  . & . & . & . & . & . & . \\
    a_{n1} & a_{n2} & . & . & . &a_{kI}& . & . & . & a_{kJ} & . & . & . & a_{nm} \\
  \end{array}
\right]^{T} \\
&=&\left[
  \begin{array}{cccccccccccccc}
    b_{11} & b_{12} & . & .& . & b_{1I} & . & . & . & b_{1J} & . & . & . & b_{1n} \\
    b_{21} & b_{22} & . & .& . & b_{2I} & . & . & . & b_{2J} & . & . & . & b_{2n} \\
    . & . & . & . & . & . & . & . & . & . & . & .& . & . \\
    . & . & . & . & . & . & . & . & . & . & . & . & .& . \\
    . & . & . & . & . & . & . & . & . & . & . & . & .& . \\
    b_{k1} & b_{k2} & . & . & . & b_{kI} & . & . & . & b_{kJ} & . & . & . & b_{kn} \\
    . & . & . & . & . & . & . & . & . & . & . & . & . & . \\
    . & . & . & . & . & . & . & . & . & . & . & . & . & . \\
    . & . & . & . & . & . & .&  . & . & . & . & . & . & . \\
    b_{n1} & b_{n2} & . & . & . &b_{kI}& . & . & . & b_{kJ} & . & . & . & b_{nn} \\
  \end{array}
\right];
\\
\Gamma(\mathscr{C}^{\ast})=M_{\mathscr{C}^{\ast}}\cdot
M^{T}_{\mathscr{C^{\ast}}}&=&\left[
  \begin{array}{cccccccccccccc}
    a_{11} & a_{12} & . & .& . & a_{1I} & . & . & . & a_{1J} & . & . & . & a_{1m} \\
    a_{21} & a_{22} & . & .& . & a_{2I} & . & . & . & a_{2J} & . & . & . & a_{2m} \\
    . & . & . & . & . & . & . & . & . & . & . & .& . & . \\
    . & . & . & . & . & . & . & . & . & . & . & . & .& . \\
    . & . & . & . & . & . & . & . & . & . & . & . & .& . \\
    a_{k1} & a_{k2} & . & . & .& 0 & . & . & . & 1 & . & . & . & a_{km} \\
    . & . & . & . & . & . & . & . & . & . & . & . & . & . \\
    . & . & . & . & . & . & . & . & . & . & . & . & . & . \\
    . & . & . & . & . & . & .&  . & . & . & . & . & . & . \\
    a_{n1} & a_{n2} & . & . & . &a_{kI}& . & . & . & a_{kJ} & . & . & . & a_{nm} \\
  \end{array}
\right]
\\
&\cdot& \left[
  \begin{array}{cccccccccccccc}
    a_{11} & a_{12} & . & .& . & a_{1I} & . & . & . & a_{1J} & . & . & . & a_{1m} \\
    a_{21} & a_{22} & . & .& . & a_{2I} & . & . & . & a_{2J} & . & . & . & a_{2m} \\
    . & . & . & . & . & . & . & . & . & . & . & .& . & . \\
    . & . & . & . & . & . & . & . & . & . & . & . & .& . \\
    . & . & . & . & . & . & . & . & . & . & . & . & .& . \\
    a_{k1} & a_{k2} & . & . & .& 0 & . & . & . & 1 & . & . & . & a_{km} \\
    . & . & . & . & . & . & . & . & . & . & . & . & . & . \\
    . & . & . & . & . & . & . & . & . & . & . & . & . & . \\
    . & . & . & . & . & . & .&  . & . & . & . & . & . & . \\
    a_{n1} & a_{n2} & . & . & . &a_{kI}& . & . & . & a_{kJ} & . & . & . & a_{nm} \\
  \end{array}
\right]^{T}
\end{eqnarray*}
\begin{eqnarray*}
&=&\left[
  \begin{array}{cccccccccccccc}
    c_{11} & c_{12} & . & .& . & c_{1I} & . & . & . & c_{1J} & . & . & . & c_{1n} \\
    c_{21} & c_{22} & . & .& . & c_{2I} & . & . & . & c_{2J} & . & . & . & c_{2n} \\
    . & . & . & . & . & . & . & . & . & . & . & .& . & . \\
    . & . & . & . & . & . & . & . & . & . & . & . & .& . \\
    . & . & . & . & . & . & . & . & . & . & . & . & .& . \\
    c_{k1} & c_{k2} & . & . & . & c_{kI} & . & . & . & c_{kJ} & . & . & . & c_{kn} \\
    . & . & . & . & . & . & . & . & . & . & . & . & . & . \\
    . & . & . & . & . & . & . & . & . & . & . & . & . & . \\
    . & . & . & . & . & . & .&  . & . & . & . & . & . & . \\
    c_{n1} & c_{n2} & . & . & . & c_{kI}& . & . & . & c_{kJ} & . & . & . & c_{nn} \\
  \end{array}
\right].
\end{eqnarray*}

An example is employed to illustrate the process of the computing of
the type-1 characteristic matrix with the proposed approach.

\begin{example} (Continuation of Example 3.3)
If we delete $x_{1}$ in $C_{1}$ and add it into $C_{3}$, then we get
$\mathscr{C}^{\ast}=\{C^{\ast}_{1},C^{\ast}_{2},C^{\ast}_{3}\}$,
where $C^{\ast}_{1}=\{x_{4}\}$, $C^{\ast}_{2}=\{x_{1},x_{2},x_{4}\}$
and $C^{\ast}_{3}=\{x_{1},x_{3},x_{4}\}$. Suppose that
$\Gamma(\mathscr{C})=(b_{ij})_{4\times 4}$ shown in Example 3.3 and
$\Gamma(\mathscr{C}^{\ast})=(c_{ij})_{4\times 4}$. Actually, we have
that $b_{ij}=c_{ij}$ for $2\leq i,j\leq 4$ and only need to compute
$c_{11}, c_{12}, c_{13} $ and $c_{14}$.
\begin{eqnarray*}
\Gamma(\mathscr{C}^{\ast})=M_{\mathscr{C}^{\ast}}\cdot
M^{T}_{\mathscr{C^{\ast}}}&=&\left[
  \begin{array}{ccc}
1 & 1 & 1 \\
0 & 1 & 0 \\
0 & 0 & 1 \\
0 & 1 & 1 \\
\end{array}
\right]\cdot \left[
  \begin{array}{cccc}
1 & 0 & 0 & 0 \\
1 & 1 & 0 & 1 \\
1 & 0 & 1 & 1 \\
\end{array}
\right]=\left[
  \begin{array}{cccc}
c_{11} & c_{12} & c_{13} & c_{14} \\
c_{21} & 1 & 0 & 1 \\
c_{31} & 0 & 1 & 1 \\
c_{41} & 1 & 1 & 1 \\
\end{array}
\right].
\end{eqnarray*}

It is easy to get that $c_{11}=c_{12}=c_{13}=c_{14}=1$ and
\begin{eqnarray*}
\Gamma(\mathscr{C}^{\ast})=\left[
  \begin{array}{cccc}
c_{11} & c_{12} & c_{13} & c_{14} \\
c_{21} & 1 & 0 & 1 \\
c_{31} & 0 & 1 & 1 \\
c_{41} & 1 & 1 & 1 \\
\end{array}
\right] &=&\left[
\begin{array}{cccc}
1 & 1 & 1 & 1 \\
1 & 1 & 0 & 1 \\
1 & 0 & 1 & 1 \\
1 & 1 & 1 & 1 \\
\end{array}
\right].
\end{eqnarray*}
\end{example}

Then, we investigate the relationship $\prod(\mathscr{C})$ and
$\prod(\mathscr{C}^{\ast})$ which are shown as follows. Concretely,
we have that $b_{ij}=c_{ij}$ for $i\neq k$ and $ j\neq k$. Thus we
only need to compute $c_{ik}$ for $1\leq i\leq n$ and $c_{kj}$ for
$1\leq j\leq n$.
\begin{eqnarray*}
\prod(\mathscr{C})=M_{\mathscr{C}}\odot M^{T}_{\mathscr{C}}&=&\left[
  \begin{array}{cccccccccccccc}
    a_{11} & a_{12} & . & .& . & a_{1I} & . & . & . & a_{1J} & . & . & . & a_{1m} \\
    a_{21} & a_{22} & . & .& . & a_{2I} & . & . & . & a_{2J} & . & . & . & a_{2m} \\
    . & . & . & . & . & . & . & . & . & . & . & .& . & . \\
    . & . & . & . & . & . & . & . & . & . & . & . & .& . \\
    . & . & . & . & . & . & . & . & . & . & . & . & .& . \\
    a_{k1} & a_{k2} & . & . & . & 1 & . & . & . & 0 & . & . & . & a_{km} \\
    . & . & . & . & . & . & . & . & . & . & . & . & . & . \\
    . & . & . & . & . & . & . & . & . & . & . & . & . & . \\
    . & . & . & . & . & . & .&  . & . & . & . & . & . & . \\
    a_{n1} & a_{n2} & . & . & . &a_{kI}& . & . & . & a_{kJ} & . & . & . & a_{nm} \\
  \end{array}
\right]\end{eqnarray*}
\begin{eqnarray*}&\odot& \left[
  \begin{array}{cccccccccccccc}
    a_{11} & a_{12} & . & .& . & a_{1I} & . & . & . & a_{1J} & . & . & . & a_{1m} \\
    a_{21} & a_{22} & . & .& . & a_{2I} & . & . & . & a_{2J} & . & . & . & a_{2m} \\
    . & . & . & . & . & . & . & . & . & . & . & .& . & . \\
    . & . & . & . & . & . & . & . & . & . & . & . & .& . \\
    . & . & . & . & . & . & . & . & . & . & . & . & .& . \\
    a_{k1} & a_{k2} & . & . & .& 1 & . & . & . & 0 & . & . & . & a_{km} \\
    . & . & . & . & . & . & . & . & . & . & . & . & . & . \\
    . & . & . & . & . & . & . & . & . & . & . & . & . & . \\
    . & . & . & . & . & . & .&  . & . & . & . & . & . & . \\
    a_{n1} & a_{n2} & . & . & . &a_{kI}& . & . & . & a_{kJ} & . & . & . & a_{nm} \\
  \end{array}
\right]^{T} \\&=&\left[
  \begin{array}{cccccccccccccc}
    b_{11} & b_{12} & . & .& . & b_{1I} & . & . & . & b_{1J} & . & . & . & b_{1n} \\
    b_{21} & b_{22} & . & .& . & b_{2I} & . & . & . & b_{2J} & . & . & . & b_{2n} \\
    . & . & . & . & . & . & . & . & . & . & . & .& . & . \\
    . & . & . & . & . & . & . & . & . & . & . & . & .& . \\
    . & . & . & . & . & . & . & . & . & . & . & . & .& . \\
    b_{k1} & b_{k2} & . & . & . & b_{kI} & . & . & . & b_{kJ} & . & . & . & b_{kn} \\
    . & . & . & . & . & . & . & . & . & . & . & . & . & . \\
    . & . & . & . & . & . & . & . & . & . & . & . & . & . \\
    . & . & . & . & . & . & .&  . & . & . & . & . & . & . \\
    b_{n1} & b_{n2} & . & . & . &b_{kI}& . & . & . & b_{kJ} & . & . & . & b_{nn} \\
  \end{array}
\right];
\\
\prod(\mathscr{C}^{\ast})=M_{\mathscr{C}^{\ast}}\odot
M^{T}_{\mathscr{C^{\ast}}}&=&\left[
  \begin{array}{cccccccccccccc}
    a_{11} & a_{12} & . & .& . & a_{1I} & . & . & . & a_{1J} & . & . & . & a_{1m} \\
    a_{21} & a_{22} & . & .& . & a_{2I} & . & . & . & a_{2J} & . & . & . & a_{2m} \\
    . & . & . & . & . & . & . & . & . & . & . & .& . & . \\
    . & . & . & . & . & . & . & . & . & . & . & . & .& . \\
    . & . & . & . & . & . & . & . & . & . & . & . & .& . \\
    a_{k1} & a_{k2} & . & . & .& 0 & . & . & . & 1 & . & . & . & a_{km} \\
    . & . & . & . & . & . & . & . & . & . & . & . & . & . \\
    . & . & . & . & . & . & . & . & . & . & . & . & . & . \\
    . & . & . & . & . & . & .&  . & . & . & . & . & . & . \\
    a_{n1} & a_{n2} & . & . & . &a_{kI}& . & . & . & a_{kJ} & . & . & . & a_{nm} \\
  \end{array}
\right]
\\
&\odot& \left[
  \begin{array}{cccccccccccccc}
    a_{11} & a_{12} & . & .& . & a_{1I} & . & . & . & a_{1J} & . & . & . & a_{1m} \\
    a_{21} & a_{22} & . & .& . & a_{2I} & . & . & . & a_{2J} & . & . & . & a_{2m} \\
    . & . & . & . & . & . & . & . & . & . & . & .& . & . \\
    . & . & . & . & . & . & . & . & . & . & . & . & .& . \\
    . & . & . & . & . & . & . & . & . & . & . & . & .& . \\
    a_{k1} & a_{k2} & . & . & .& 0 & . & . & . & 1 & . & . & . & a_{km} \\
    . & . & . & . & . & . & . & . & . & . & . & . & . & . \\
    . & . & . & . & . & . & . & . & . & . & . & . & . & . \\
    . & . & . & . & . & . & .&  . & . & . & . & . & . & . \\
    a_{n1} & a_{n2} & . & . & . &a_{kI}& . & . & . & a_{kJ} & . & . & . & a_{nm} \\
  \end{array}
\right]^{T}
\\
&=&\left[
  \begin{array}{cccccccccccccc}
    c_{11} & c_{12} & . & .& . & c_{1I} & . & . & . & c_{1J} & . & . & . & c_{1n} \\
    c_{21} & c_{22} & . & .& . & c_{2I} & . & . & . & c_{2J} & . & . & . & c_{2n} \\
    . & . & . & . & . & . & . & . & . & . & . & .& . & . \\
    . & . & . & . & . & . & . & . & . & . & . & . & .& . \\
    . & . & . & . & . & . & . & . & . & . & . & . & .& . \\
    c_{k1} & c_{k2} & . & . & . & c_{kI} & . & . & . & c_{kJ} & . & . & . & c_{kn} \\
    . & . & . & . & . & . & . & . & . & . & . & . & . & . \\
    . & . & . & . & . & . & . & . & . & . & . & . & . & . \\
    . & . & . & . & . & . & .&  . & . & . & . & . & . & . \\
    c_{n1} & c_{n2} & . & . & . & c_{kI}& . & . & . & c_{kJ} & . & . & . & c_{nn} \\
  \end{array}
\right].
\end{eqnarray*}

We employ an example to illustrate the process of the computing of
the type-2 characteristic matrix with the proposed approach.

\begin{example} (Continuation of Example 4.12)
Suppose that $\prod(\mathscr{C})=(b_{ij})_{4\times 4}$ shown in
Example 3.3 and $\prod(\mathscr{C}^{\ast})=(c_{ij})_{4\times 4}$.
Actually, we have that $b_{ij}=c_{ij}$ for $2\leq i,j\leq 4$. Thus
we only need to compute $c_{11}, c_{12}, c_{13}, c_{14},c_{21},
c_{31}$ and $ c_{41}$.
\begin{eqnarray*}
\prod(\mathscr{C}^{\ast})=M_{\mathscr{C}^{\ast}}\odot
M^{T}_{\mathscr{C^{\ast}}}&=&\left[
  \begin{array}{ccc}
1 & 1 & 1 \\
0 & 1 & 0 \\
0 & 0 & 1 \\
0 & 1 & 1 \\
\end{array}
\right]\odot \left[
  \begin{array}{cccc}
1 & 0 & 0 & 0 \\
1 & 1 & 0 & 1 \\
1 & 0 & 1 & 1 \\
\end{array}
\right]=\left[
  \begin{array}{cccc}
c_{11} & c_{12} & c_{13} & c_{14} \\
c_{21} & 1 & 0 & 1 \\
c_{31} & 0 & 1 & 1 \\
c_{41} & 0 & 0 & 1 \\
\end{array}
\right].
\end{eqnarray*}
Consequently, we get that $c_{11}=c_{21}=c_{31}=c_{41}=1,
c_{12}=c_{13}=c_{14}=0 $. Then we have that
\begin{eqnarray*}
\prod(\mathscr{C}^{\ast})=\left[
  \begin{array}{cccc}
1 & 0 & 0 & 0 \\
1 & 1 & 0 & 1 \\
1 & 0 & 1 & 1 \\
1 & 0 & 0 & 1 \\
\end{array}
\right].
\end{eqnarray*}
\end{example}

Due to the change of the attribute value of the object $x_{k}\in U$,
maybe we have that $x$ does not belong to any element of
$\mathscr{C}$. So we suppose that $C^{\ast}_{m+1}=\{x_{k}\}$ and
$\mathscr{C}^{\ast}=\{C^{\ast}_{1},C^{\ast}_{2},...,C^{\ast}_{m},C^{\ast}_{m+1}\}$.
We investigate the relationship $M_{\mathscr{C}}$ and
$M_{\mathscr{C}^{\ast}}$ and suppose that
$\Gamma(\mathscr{C})=(b_{ij})_{n\times n}$,
$\Gamma(\mathscr{C}^{\ast})=(c_{ij})_{n\times n}$,
$\prod(\mathscr{C})=(b^{\ast}_{ij})_{n\times n}$ and
$\prod(\mathscr{C}^{\ast})=(c^{\ast}_{ij})_{n\times n}$. Concretely,
we have that $b_{ij}=c_{ij}$ and $b^{\ast}_{ij}=c^{\ast}_{ij}$ for
$i\neq k$ and $ j\neq k$. Thus, we can obtain
$\Gamma(\mathscr{C}^{\ast})$ and $\prod(\mathscr{C}^{\ast})$ by
computing $c_{ik}$ ($1\leq i\leq n$), $c^{\ast}_{ik}$ ($1\leq i\leq
n$) and $c^{\ast}_{kj}$ ($1\leq j\leq n$).

The time complexity of computing the type-1 (respectively, type-2)
characteristic matrix is $\mathcal {O}(n)+\mathcal {O}(n^{2})$ if
$|U^{\ast}|=n$, $|\mathscr{C}|=m$ and an attribute value is revised.
But the time complexity is $\mathcal {O}(m\ast n^{2})$ by using the
concept of the type-1 (respectively, type-2) characteristic matrix.
Therefore, we can get the type-1 and type-2 characteristic matrixes
with less time by using the proposed approach.

\section{Conclusions}

The approximations of concepts and feature selections in the dynamic
information system are important works of the rough set theory, and
they are challenging issues in the field of the artificial
intelligence. In this paper, we have introduced two approaches to
constructing the approximations of concepts in the covering
approximation space. After that, we have constructed the
characteristic matrixes of five types of the dynamic coverings with
the proposed approach. Additionally, we have employed several
examples to illustrate the process of computing the characteristic
matrixes of the dynamic coverings by using an incremental approach.

In the future, we will propose more effective approaches to
constructing the characteristic matrixes of the covering.
Additionally, we will focus on the development of effective
approaches for knowledge discovery of dynamic information systems.

\section*{ Acknowledgments}

We would like to thank the anonymous reviewers very much for their
professional comments and valuable suggestions. This work is
supported by the National Natural Science Foundation of China (NO.
11071061) and the National Basic Research Program of China (NO.
2010CB334706, 2011CB311808).

\end{document}